\documentclass[aps,prd,twocolumn,superscriptaddress,preprintnumbers,nofootinbib,10pt]{revtex4-1}
\pdfoutput=1

\usepackage{graphicx}
\usepackage{mathrsfs}
\usepackage{amsmath,amstext,amsfonts,amssymb}
\usepackage{color}

\definecolor{darkblue}{rgb}{0,0,0.5}
\usepackage[colorlinks,linkcolor=darkblue,citecolor=darkblue,urlcolor=darkblue]{hyperref}

\allowdisplaybreaks

\newcommand{\beq}{\begin{equation}}
\newcommand{\eeq}{\end{equation}}
\newcommand{\bea}{\begin{eqnarray}}
\newcommand{\eea}{\end{eqnarray}}
\newcommand{\ra}{\rightarrow}
\newcommand{\eps}{\epsilon}
\newcommand{\mh}{M_h}

\begin{document}

\preprint{FERMILAB-PUB-12-550-T}
\preprint{EFI-12-25}
\preprint{ANL-HEP-PR-12-75}

\title{A Minimal Flavor Violating 2HDM at the LHC}

\author{Wolfgang~Altmannshofer}
\affiliation{Fermi National Accelerator Laboratory, P.O.~Box 500, Batavia, IL 60510, USA}

\author{Stefania~Gori}
\affiliation{Enrico Fermi Institute, University of Chicago, Chicago, IL 60637, USA}
\affiliation{HEP Division, Argonne National Laboratory, 9700 Cass Ave., Argonne, IL 60439}

\author{Graham~D.~Kribs}
\affiliation{Department of Physics, University of Oregon, Eugene, OR 97403, USA}

\begin{abstract}

We explore the phenomenology of a two Higgs doublet model where 
both Higgs doublets couple to up-type and down-type fermions 
with couplings determined by the minimal flavor violation ansatz.  
This ``2HDM Type MFV'' generalizes 2HDM Types I-IV, 
where the decay rates of $h \ra b\bar{b}$ and $h \ra \tau^+\tau^-$ 
are governed by MFV couplings independent of the Higgs couplings to 
gauge bosons or the top quark. 
To determine the implications of the present Higgs data on the model, 
we have performed global fits to all relevant data.
Several surprisingly large effects on the light Higgs phenomenology 
can arise:
(1) The modified couplings of the Higgs to fermions can enhance the 
$h \ra \gamma\gamma$ rate significantly in both VBF production 
(up to a factor of $3$ or more) and the inclusive rate 
(up to a factor of $1.5$ or more).  
(2) In the 2HDM Type MFV, the constraints on a light charged Higgs 
are milder than in 2HDM Types I-IV\@.  Thus, there can be substantial 
charged Higgs loop contribution to the di-photon rate, 
allowing further enhancements of the di-photon rates.
(3) The $h \ra \tau^+\tau^-$ rate can be (highly) suppressed, 
independently of the other decay channels. Furthermore, we studied 
the correlation between the light Higgs and the heavy Higgs phenomenology. 
We showed that even small deviations from the decoupling limit would 
imply good prospects for the detection of the heavy Higgs boson.
In some regions of parameter space, a substantial range of 
$M_H$ is already either ruled out or on the edge of detection. 
Finally we investigated the possibility that the heavy Higgs is
close in mass to the light Higgs, providing additional 
$h/H \ra b\bar{b}$ rate, as well as confounding the extraction 
of properties of the Higgs bosons.

\end{abstract}

\maketitle

\section{Introduction}

How many scalar Higgs doublets are in Nature?
The intriguing possibility of additional Higgs doublets has taken
on heightened importance in light of the observations by 
ATLAS~\cite{:2012gk} and CMS~\cite{:2012gu} of a particle consistent 
with the Higgs boson.  One of the simplest extensions of the 
Higgs sector of the standard model (SM) is a two-Higgs doublet model (2HDM).  
Glashow and Weinberg long ago recognized that a general 2HDM, in which 
both Higgs doublets couple arbitrarily to the quarks and leptons 
of the standard model, would induce excessively large flavor-changing 
neutral currents (FCNCs) \cite{Glashow:1976nt}.  This has led to the 
various well-known (and less-known) classifications or 
``Types'' of 2HDM models with natural flavor conservation, 
where the different types of fermions (up-type quarks, down-type quarks 
and leptons) couple to one Higgs doublet only
(for a recent review see~\cite{Branco:2011iw}).
We will refer to these models as 2HDM Types I-IV, 
following~\cite{Craig:2012vn}.

In the absence of enlarged symmetries, there is nothing to enforce 
exact flavor-conservation of the Higgs couplings. 
Thus, it behooves us to understand the full extent 
of what is possible within a general 2HDM, when 
both Higgs doublets couple to all quarks and leptons, 
but with controlled couplings such that FCNC constraints 
are satisfied.  
Various mechanisms have been discussed in the literature to protect 2HDMs from too large FCNCs~\cite{D'Ambrosio:2002ex,Chivukula:1987py,Hall:1990ac,Buras:2000dm,Branco:1996bq,DiazCruz:2004tr,Pich:2009sp,Botella:2009pq,Gupta:2009wn,Botella:2011ne,Mrazek:2011iu}. Among them, minimal flavor violation (MFV)~\cite{D'Ambrosio:2002ex,Chivukula:1987py,Hall:1990ac,Buras:2000dm} provides a simple 
ansatz to allow general Higgs couplings without excessive FCNCs.

In this paper we consider a 2HDM in which both Higgs
doublets couple to quarks and leptons with MFV
couplings.  This model was proposed in \cite{D'Ambrosio:2002ex}
and it has been studied in~\cite{Mantry:2007ar,Buras:2010mh,Buras:2010zm,Trott:2010iz,Blankenburg:2011ca,Cline:2011mm,Altmannshofer:2011iv}, where the main focus was on the flavor phenomenology of the model. In this work we study the model in view of the current results from Higgs searches.
As we will see, the 2HDM Type MFV generalizes the various
2HDM Types I-IV that have been recently considered in 
the post-125 GeV literature~\cite{Ferreira:2011aa,Burdman:2011ki,Ferreira:2012my,Blum:2012kn,Azatov:2012wq,Craig:2012vn,Alves:2012ez,Swiezewska:2012ej,Craig:2012pu}.  
The principle differences between the various Types of 2HDM are 
the couplings to the $b$ quark and the $\tau$ lepton.
In the 2HDM Type MFV, these couplings can be modified independently 
from each other, and also independently from the Higgs couplings 
to the top quark and the weak gauge bosons. 
Those aspects of Higgs physics dependent on the couplings to 
the top quark or gauge bosons are largely equivalent to the other
Types of 2HDM\@.  Hence, the production cross section through 
gluon fusion, vector boson fusion (VBF) production, and associated production 
can be fairly accurately described by 2HDM Types I-IV couplings
(determined by $\tan\beta$ and $\alpha$).  
By contrast, the 2HDM Type MFV can have completely different widths 
$\Gamma_{b\bar{b}}$ and $\Gamma_{\tau\tau}$ as compared with any of the 
flavor-preserving Types of 2HDMs\@.  This has very significant 
effects on Higgs physics, not only on the branching fractions
$h \to b\bar{b}$ and $h \to \tau^+\tau^-$, but to all modes since 
the total width of the light Higgs boson is 
dominated by $\Gamma_{b\bar{b}}$.

There is one additional intriguing possibility, in which 
both light and heavy CP-even Higgs bosons are light
(for a recent discussion, see 
\cite{Cervero:2012cx,Gunion:2012gc,Batell:2012mj,Gunion:2012he,Belanger:2012he}).
The presence of two light CP-even Higgs bosons with 
modified couplings to $b\bar{b}$ and $\tau^+\tau^-$ 
can potentially allow for increased branching fractions
to multiple interesting modes.  For instance, the $\gamma\gamma$ rate
can be substantially larger through one light Higgs boson 
without sacrificing a significant rate into $b\bar{b}$ 
due to the presence of the second light Higgs boson.  
The hints for rates into $b\bar{b}$ by the CDF and D0 
collaborations~\cite{:2012cn} could, if this scenario is right, 
suggest they have observed the second Higgs boson.

\section{Higgs Couplings to Fermions}\label{sec:coupl_Higgs_fermion}

The most general couplings of two Higgs doublets $H_1$ and $H_2$,
with hypercharge $1/2$ and $-1/2$ respectively,
to the SM fermions have the form
\begin{eqnarray}
\mathcal{L} &\supset& \phantom{+} (y_u)_{ij} ~H_2 \bar Q_i U_j + (\tilde y_u)_{ij} ~H_1^\dagger \bar Q_i U_j  \\
&& + (y_d)_{ij} ~H_1 \bar Q_i D_j + (\tilde y_d)_{ij} ~H_2^\dagger \bar Q_i D_j \nonumber \\
&& + (y_\ell)_{ij} ~H_1 \bar L_i E_j + (\tilde y_\ell)_{ij} ~H_2^\dagger \bar L_i E_j  ~~+ {\rm h.c.}~, \nonumber
\end{eqnarray}
where both Higgs bosons couple to up-type and down-type fermions 
as well as leptons.  The resulting masses of the fermions are given by
\begin{eqnarray}
m_u &=& \frac{v s_\beta}{\sqrt{2}} \Big(y_u + \frac{1}{t_\beta} \tilde y_u \Big)~,~~ m_d = \frac{v c_\beta}{\sqrt{2}} \Big( y_d + t_\beta \tilde y_d \Big) ~, \nonumber \\
m_\ell &=& \frac{v c_\beta}{\sqrt{2}} \Big( y_\ell + t_\beta \tilde y_\ell \Big) ~,
\end{eqnarray}
where $v_2 = v s_\beta$ and $v_1 = v c_\beta$ are the $SU(2)_L$ 
breaking vacuum expectation values (vevs) of the two doublets, 
$\tan\beta = t_\beta = v_2 / v_1$ is their ratio, 
and $v \simeq 246$~GeV\@.

In general, one cannot take generic $\tilde{y}$ and $y$ couplings
without inducing huge tree-level contributions to FCNCs.  
One way to avoid tree-level FCNCs in a 2HDM is to impose an approximate 
discrete or continuous symmetry to the model, 
where Yukawa couplings to only one 
Higgs doublet are allowed at tree-level.  Then, small couplings to 
the other Higgs doublet can arise once loop effects or 
higher dimensional operators 
are considered~\cite{Buras:2010mh}.\footnote{Well 
known examples are Minimal Supersymmetric Standard Model 
(MSSM) where the $\tilde y_i$ are loop induced~\cite{Hempfling:1993kv,Hall:1993gn,Carena:1994bv,Hamzaoui:1998nu}, or beyond the MSSM (BMSSM) 
models, where they can arise from higher dimensional operators~\cite{Dine:2007xi,Antoniadis:2008es,Altmannshofer:2011iv}.}
These already small couplings can induce sizable contributions to FCNCs.
We ensure that FCNCs are avoided by taking 
the couplings $\tilde{y}$ to obey the MFV 
ansatz~\cite{D'Ambrosio:2002ex,Chivukula:1987py,Hall:1990ac,Buras:2000dm}. 
In the quark sector, the MFV assumption states that there are only 
two spurions that break the global $SU(3)^3$ quark flavor symmetry 
of the standard model gauge sector. This implies that the couplings 
$\tilde y$ and $y$ are not independent of each other, allowing us
to write~\cite{D'Ambrosio:2002ex}
\begin{eqnarray} \label{eq:yutilde}
\tilde y_u &=& \epsilon_u y_u 
               + \epsilon_u^\prime y_u y_u^\dagger y_u 
               + \epsilon_u^{\prime\prime} y_d y_d^\dagger y_u + \dots ~, \\ 
\label{eq:ydtilde}
\tilde y_d &=& \epsilon_d y_d 
               + \epsilon_d^\prime y_d y_d^\dagger y_d 
               + \epsilon_d^{\prime\prime} y_u y_u^\dagger y_d + \dots ~,
\end{eqnarray}
with parameters $\epsilon_i$ that can in general be complex.
The terms with $\epsilon_i^{\prime\prime}$ as well as the other 
higher order terms containing both $y_u$ and $y_d$ couplings 
lead to flavor changing neutral Higgs couplings, 
that are nevertheless controlled by CKM matrix elements and 
thus naturally small. These terms can lead to interesting effects 
in $B$ physics as discussed in~\cite{Buras:2010mh,Buras:2010zm,Trott:2010iz,Cline:2011mm,Altmannshofer:2011iv}.
Here we are interested in the impact of the flavor conserving part 
of the Higgs couplings on Higgs collider phenomenology. 
We remark that the higher order terms in the expansions 
in Eqs.~(\ref{eq:yutilde}-\ref{eq:ydtilde}) 
can induce non-universalities between the couplings 
to the first two and the third generation of fermions. 
Since Higgs phenomenology is dominated by the third generation, 
we will set all the higher order terms to 
zero.  (In fact, even if the higher order terms were generated radiatively
\cite{Braeuninger:2010td}, they do not impact Higgs phenomenology.)  
We consider only 
\begin{equation}
\tilde y_u = \epsilon_u y_u ~,~~ \tilde y_d = \epsilon_d y_d ~.
\end{equation}
This is the ``aligned 2HDM'' framework presented in~\cite{Pich:2009sp}.
In the following we concentrate on the CP conserving case with 
$\epsilon_q$ real.  For the lepton sector, we analogously assume 
that 
\begin{equation}
\tilde y_\ell = \epsilon_\ell y_\ell ~,
\end{equation}
with a real proportionality factor $\epsilon_\ell$, leading to
 flavor-conserving Higgs-lepton couplings 
(by contrast, see~\cite{Blankenburg:2012ex,Harnik:2012pb} 
for studies of lepton flavor-violating Higgs decays).
The $\epsilon_i$ are flavor-universal, so we can interchangeably 
use the subscripts $u~\leftrightarrow~t$, $d~\leftrightarrow~b$ and 
$\ell~\leftrightarrow~\tau$ in the following.

In the setup outlined above, we can directly express the Yukawa couplings 
in terms of the measured quark masses
\begin{eqnarray}
y_u &=& \frac{\sqrt 2 m_u}{v s_\beta} \frac{1}{1 + \epsilon_u / t_\beta} ~,~~ 
  y_d = \frac{\sqrt 2 m_d}{v s_\beta} \frac{t_\beta}{1 + \epsilon_d t_\beta} ~, 
  \nonumber \\
y_\ell &=& \frac{\sqrt 2 m_\ell}{v s_\beta} \frac{t_\beta}{1 + \epsilon_\ell t_\beta} ~.
\end{eqnarray}
In general, the three real $\epsilon_i$ parameters are 
not all physical. 
One of them can always be reabsorbed by a redefinition of the 
two original Higgs doublets.  For example, we can choose
$H_u \propto H_2 + \epsilon_u H_1^\dagger$ to be the combination 
of Higgs fields that couples to up-type quarks.  
This choice of basis for the Higgs doublets corresponds 
to setting $\epsilon_u = 0$, which we assume, without loss of generality, 
throughout the paper.  This implies the 2HDM Type MFV 
couplings to the top quark are identical to the other 2HDM Types I-IV\@. 
Furthermore, this choice of basis uniquely defines $\tan\beta$ 
as the ratio of the vev of the $H_u$ field with coupling $\epsilon_u=0$ 
and the vev of the orthogonal Higgs field $H_d$ 
(for a discussion of basis invariant quantities in 2HDMs 
see~\cite{Davidson:2005cw,Ferreira:2010jy,Haber:2010bw}).

We now write the couplings of the Higgs boson mass eigenstates 
with the SM quarks and leptons, as well as the gauge bosons.
The two Higgs doublets comprise 8 real scalar fields, 
three of which are the usual Goldstone bosons $G$ and $G^\pm$ that provide 
the longitudinal components of the $Z$ and $W^\pm$ bosons. The remaining 
physical Higgs bosons consist of two CP-even scalars $h$ and $H$, 
one CP-odd scalar $A$ and the charged Higgs $H^\pm$.  
Their interaction Lagrangian is
\begin{eqnarray}
\mathcal{L}_\text{int} &=&{} 
   - \frac{m_{u_i}}{v} ~ \bar u_i P_R u_i ~ 
     \left( iG +i \xi_u^A A + \xi_u^H H + \xi_u^h h \right) 
     \nonumber \\
&&{} - \frac{m_{d_i}}{v} ~ \bar d_i P_R d_i ~ 
     \left( -iG +i \xi_d^A A + \xi_d^H H + \xi_d^h h \right) 
     \nonumber \\
&&{} - \frac{m_{\ell_i}}{v} ~ \bar \ell_i P_R \ell_i ~ 
     \left( -iG +i \xi_\ell^A A + \xi_\ell^H H + \xi_\ell^h h \right) 
     \nonumber \\
&&{} - \sqrt{2} \frac{m_{d_j}}{v} ~V_{ij}~ \bar u_i P_R d_j ~ 
     \left( G^+ + \xi_d^+ H^+ \right) 
     \nonumber \\
&&{} - \sqrt{2} \frac{m_{u_j}}{v} ~V_{ij}^*~ \bar d_i P_R u_j ~ 
     \left( G^- + \xi_u^- H^- \right) 
     \nonumber \\
&&{} - \sqrt{2} \frac{m_{\ell_i}}{v} ~ \bar \nu_i P_R \ell_i ~ 
     \left( G^+ + \xi_\ell^+ H^+ \right) 
     \nonumber \\
&&{} + \frac{g_2^2}{2} v \left[ \frac{Z^2}{2c_W^2} + W^2 \right] 
     \left( \xi_V^h \, h + \xi_V^H \, H\right) 
     \, ,
\end{eqnarray}
where $V_{ij}$ are elements of the CKM matrix and $c_W$ 
is the cosine of the Weinberg angle.
For our purposes it is justified to take the neutrinos to be massless.  
For the reduced couplings $\xi$, 
that parameterize the deviations of the Higgs couplings from 
the Yukawa couplings in the SM, one finds
\begin{eqnarray} \label{eq:reduced_couplings}
\xi_u^h &=& 
  \frac{c_\alpha}{s_\beta} ~,~~ 
\xi_u^H = 
  \frac{s_\alpha}{s_\beta} ~, 
  \nonumber \\
\xi_u^A = \xi_u^- &=& 
  \frac{1}{t_\beta} ~, 
  \nonumber \\
\xi_d^h &=& 
  \frac{-s_\alpha + \epsilon_d c_\alpha}{c_\beta + \epsilon_d s_\beta} ~,~~ 
\xi_d^H =
   \frac{c_\alpha + \epsilon_d s_\alpha}{c_\beta + \epsilon_d s_\beta} ~,
  \nonumber \\
\xi_d^A = \xi_d^- &=& 
  \frac{t_\beta -\epsilon_d}{1 + \epsilon_d t_\beta}  ~, 
  \nonumber \\
\xi_\ell^h &=& 
  \frac{-s_\alpha + \epsilon_\ell c_\alpha}{c_\beta + \epsilon_\ell s_\beta} ~,~~ 
\xi_\ell^H = 
  \frac{c_\alpha + \epsilon_\ell s_\alpha}{c_\beta + \epsilon_\ell s_\beta} ~, 
  \nonumber \\
\xi_\ell^A = \xi_\ell^- &=& 
  \frac{t_\beta -\epsilon_\ell}{1 + \epsilon_\ell t_\beta} ~, 
  \nonumber \\
\xi_V^h &=& s_{\beta - \alpha}\,,~~ \xi_V^H =  c_{\beta - \alpha} ~.
\end{eqnarray}
The angle $\alpha$ diagonalizes the mass matrix of the two CP-even Higgs bosons. 
The couplings of the Higgs bosons to the top quark and the gauge bosons 
are identical to the 2HDM Types I-IV\@.  The couplings to the
down-type quarks and leptons are in general different, parameterized
by $\epsilon_d$ and $\epsilon_\ell$.  These couplings interpolate 
continuously between the couplings of the well-studied 2HDM Types I-IV, 
recovering the different Types in the following limits:
\begin{eqnarray}
\eps_d \ra \infty, \; \eps_\ell \ra \infty  &\qquad&  \mbox{(Type I)} 
  \nonumber \\
\eps_d \ra 0, \;   \eps_\ell \ra 0         &\qquad&  \mbox{(Type II)} 
  \nonumber \\
\eps_d \ra \infty, \; \eps_\ell \ra 0       &\qquad&  \mbox{(Type III)} 
  \nonumber \\
\eps_d \ra 0, \;    \eps_\ell \ra \infty    &\qquad&  \mbox{(Type IV)} \, .
\label{eq:2hdmcouplinglimits}
\end{eqnarray}
The Higgs couplings satisfy the following sum rules
\begin{eqnarray} \label{eq:sumrule_u}
1 + (\xi_u^A)^2 &=& (\xi_u^h)^2 + (\xi_u^H)^2 \nonumber \\
&=& 1+ \frac{1}{t_\beta^2} ~, \\
1 + (\xi_d^A)^2 &=& (\xi_d^h)^2 + (\xi_d^H)^2 \nonumber \\
&=& \left( 1+ t_\beta^2 \right) \frac{1 + \epsilon_d^2}{(1+\epsilon_d t_\beta)^2} ~, \\\label{eq:sumtau}
1 + (\xi_\ell^A)^2 &=& (\xi_\ell^h)^2 + (\xi_\ell^H)^2 \nonumber \\
&=& \left( 1+ t_\beta^2 \right) \frac{1 + \epsilon_\ell^2}{(1+\epsilon_\ell t_\beta)^2} ~, \\ \label{eq:sumrule_V}
1 &=& (\xi_V^h)^2+(\xi_V^H)^2 ~.
\end{eqnarray}

We complete this section with a comment on 
the free parameters of the model and on the Higgs potential:
Crucial parameters for our analysis are $\epsilon_d$ and $\epsilon_\ell$, 
the angles $\alpha$ and $\beta$ and also the masses of the 
physical Higgs bosons $M_h$, $M_H$, $M_A$, and $M_{H^\pm}$. 
The masses and angles are determined by the parameters of the Higgs potential.
As presented in Appendix~\ref{sec:appa}, we consider the most general 2HDM 
scalar potential, including also quartic couplings that are often 
not considered in the literature, e.g., 
$(H_2 H_1)^2$, $(H_2 H_1)H_1^\dagger H_1$, 
and $(H_2 H_1)H_2^\dagger H_2$, c.f.~Eq.~(\ref{eq:Higgs_potential}). 
Indeed, any symmetry imposed to forbid all of these operators would also 
forbid the fermionic couplings proportional to $\epsilon$ to the
other Higgs doublet. 
Given the most general structure of the potential allows us to 
treat the Higgs masses $M_h$, $M_H$, $M_A$, and $M_{H^\pm}$, 
as well as the angles $\alpha$ and $\beta$, as free parameters in 
our numerical analysis of Secs.~\ref{sec:h} and~\ref{sec:hH}. 
This approach is justified so long as we are \emph{not} in the 
decoupling regime $M_A^2 \gg v^2$, where the Higgs masses and 
the angles are strongly correlated:
\begin{equation}
M_A^2 = M_H^2 + O(\lambda_i v^2) = M_{H^\pm}^2 + O(\lambda_i v^2) ~,
\end{equation}
\begin{equation}
\alpha = \beta - \pi/2 + O(\lambda_i v^2/M_A^2) ~.
\end{equation}
We explicitly checked that the scenarios discussed below can be 
realized by appropriate choices of the quartic couplings in the 
Higgs potential that are compatible with constraints from 
perturbativity and vacuum stability.

\section{Higgs Production Cross Sections and Branching Ratios}

For the production and decay of the light Higgs boson, there are 
several changes with respect to the standard model rates.  Some of these
changes do \emph{not} depend on the Type of the 2HDM\@.
This is because, as we saw from the previous section, 
the modification of the top couplings 
$\xi_u^{h}$
as well as the modifications of the 
gauge boson couplings $\xi_V^{h}$ are the same with
respect to the various Types of 2HDMs.  This implies several 
simplifications when comparing one Type of 2HDM with another.
Here we wish to present the cross sections in the 2HDM Type MFV, 
as well as to elucidate the similarities or differences between
the different Types of 2HDMs.

\subsection{Production Cross Sections}

At the LHC and the Tevatron, the dominant production mechanism for a 
SM-like Higgs is gluon fusion. The modifications of the couplings 
of the Higgs to top and bottom quarks result in a modification of this 
production cross section, that proceeds through top and bottom loops. 
In our numerical analysis we use \verb|HIGLU|~\cite{Spira:1995mt} 
to compute the various parts of the gluon fusion cross section
including top and bottom quarks.
Numerically, the bottom quark loop contribution is generally 
quite small.  In the SM, it contributes at the level of $\sim$10\% 
for a Higgs boson with $M_h = 125$~GeV\@. 
This also follows for 2HDM models because, 
as we will see in the next section, the ATLAS and CMS data do 
not favor a large enhancement of the Higgs-bottom coupling.
If we consider only the top quark contribution, 
then using the results from the previous section, 
the modification of the top coupling $\xi_u^h = c_\alpha/s_\beta$ 
is the same among the various Types of 2HDMs.  
Hence, the result for the gluon fusion 
cross section takes a simple (approximate) form 
\begin{eqnarray}\label{eq:sigmagg}
\sigma^{\rm MFV}_{ggh} &\simeq& \sigma^{\rm 2HDM}_{ggh} 
   \; \simeq \; \sigma^{\rm SM}_{ggh, tt} \times (\xi_u^h)^2 
\end{eqnarray}
where $\sigma^{\rm MFV}_{ggh}$ is the cross section in the the 2HDM Type MFV,
$\sigma^{\rm 2HDM}_{ggh}$ is the cross section in the the 2HDM Types I-IV,
and $\sigma^{\rm SM}_{ggh, tt}$ is the top loop contribution 
of the standard model.  

Higgs production through vector boson fusion or in association 
with $W$ or $Z$ bosons plays an important role in Higgs searches. 
In 2HDMs, they scale to an excellent approximation 
with the coupling of the Higgs boson to weak vector bosons
\begin{equation}
\frac{\sigma^{\rm MFV}_{\rm VBF}}{\sigma_{\rm VBF}^{\rm SM}} = 
  \frac{\sigma^{\rm MFV}_{Wh}}{\sigma_{Wh}^{\rm SM}} = 
  \frac{\sigma^{\rm MFV}_{Zh}}{\sigma_{Zh}^{\rm SM}} = (\xi_V^h)^2 ~,
\end{equation}
and thus
\begin{equation}
\frac{\sigma^{\rm MFV}_{\rm VBF}}{\sigma_{\rm VBF}^{\rm 2HDM}} = 
  \frac{\sigma^{\rm MFV}_{Wh}}{\sigma_{Wh}^{\rm 2HDM}} = 
  \frac{\sigma^{\rm MFV}_{Zh}}{\sigma_{Zh}^{\rm 2HDM}} = 1 ~.
\end{equation}
In our numerical results, 
we take the SM cross sections from the LHC Higgs cross section working 
group~\cite{Dittmaier:2011ti,Dittmaier:2012vm}.

Finally, Higgs boson production in association with top or bottom quarks 
is strongly suppressed in the SM, but can be important if the 
corresponding Higgs couplings are enhanced.  
For Higgs production in association with top quarks, we find 
\begin{eqnarray}
\sigma^{\rm MFV}_{tth} &=& \sigma^{\rm 2HDM}_{tth} \; \simeq \; 
  \sigma_{tth}^{\rm SM} \times (\xi_u^h)^2 ~.
\end{eqnarray}
Again, since the 2HDM Type MFV model shares the same modified 
top quark coupling as the 2HDM Types I-IV, they lead to
a modified cross section that is invariant with respect to the Type
of model.  In our numerical results, we take $\sigma_{tth}^{\rm SM}$ 
from~\cite{Dittmaier:2011ti,Dittmaier:2012vm}.

Finally, we comment on the production cross section of the heavy 
CP-even Higgs $H$ and the CP-odd Higgs $A$.  For large couplings 
to bottom quarks, the main production mode is in association 
with bottom quarks.  Here the cross section is in general 
completely different compared to 2HDM Types I-IV,
\begin{eqnarray}
\sigma^{\rm MFV}_{bbH,A} &\simeq& 
    \sigma_{bbH,A}^{\rm SM} \times (\xi_b^{H,A})^2 ~.
\end{eqnarray}
In our numerical results, 
we use \verb|bbh@nnlo|~\cite{Harlander:2003ai} to compute the 
SM cross section $\sigma_{bbH}^{\rm SM}$. The theoretical 
uncertainties of all SM production cross sections are also taken 
from~\cite{Dittmaier:2011ti,Dittmaier:2012vm}.

In summary, much of the light Higgs boson production cross sections 
in the 2HDM Type MFV are unchanged with respect to a 2HDM Type I-IV\@,
with the notable exception of the production of the heavy Higgs bosons 
in association with bottom quarks.

\subsection{Decay Rates and Branching Ratios}

The partial widths of the Higgs bosons into fermions 
and weak gauge bosons can be written as
\begin{eqnarray}
\Gamma_{hff} &\simeq& \Gamma_{hff}^{\rm SM} \times (\xi_f^h)^2 ~,\\
\Gamma_{hVV} &\simeq& \Gamma_{hVV}^{\rm SM} \times (\xi_V^h)^2 ~,
\end{eqnarray}
where $\Gamma_i^{\rm SM}$ are the corresponding decay width of the 
SM Higgs boson. In our numerical analysis we compute these 
SM decay widths using \verb|HDECAY|~\cite{Djouadi:1997yw}.
For the decays of the Higgs into gluons and photons we \emph{define} the
effective couplings
\begin{eqnarray} \label{eq:hgaga}
\Gamma_{h\gamma\gamma} &=& 
  \Gamma_{h\gamma\gamma}^{\rm SM} \times (\xi_\gamma^h)^2 ~, \\ \label{eq:hgg}
\Gamma_{hgg} &=& \Gamma_{hgg}^{\rm SM} \times (\xi_g^h)^2 ~,
\end{eqnarray}
and compute at leading order
\begin{equation}
(\xi_\gamma^h)^2 =
\frac{\Gamma_{h\gamma\gamma}^{\rm LO}}{\Gamma_{h\gamma\gamma}^{\rm SM,LO}} ~,~~
(\xi_g^h)^2 = \frac{\Gamma_{hgg}^{\rm LO}}{\Gamma_{hgg}^{\rm SM,LO}}~.
\end{equation}
As we only compute the ratio of partial widths, higher order corrections 
are expected to be small. To obtain absolute values of the decay widths 
into gluons and photons we use $\Gamma_i^{\rm SM}$ from 
\verb|HDECAY|~\cite{Djouadi:1997yw} in Eqs.~(\ref{eq:hgaga}) 
and~(\ref{eq:hgg}).

Some of the lightest Higgs partial widths are Type-invariant, 
while others are Type-dependent.
The Type-invariant widths include\footnote{As already explained 
for the gluon fusion production cross section in Eq.~(\ref{eq:sigmagg}), 
contributions from bottom loops can be neglected to a first approximation.}
\begin{equation}
\frac{\Gamma^{\rm MFV}_{VV}}{\Gamma^{\rm 2HDM}_{VV}} = 1 ~,~~
\frac{\Gamma^{\rm MFV}_{gg}}{\Gamma^{\rm 2HDM}_{gg}} \simeq 1 \, .
\end{equation}
At leading order, the width into diphotons depends on the
Higgs coupling to the $W$ boson as well as the top quark, 
which are both Type-independent couplings.  In addition, 
all Types of 2HDMs can have a charged Higgs contribution to the 
diphoton and $Z\gamma$ decay amplitudes.  
The size of the charged Higgs contribution depends on the
the scalar trilinear coupling $\lambda_{h H^\pm H^\pm}$ 
and the charged Higgs mass.  If the coupling and charged
Higgs mass are the same between two different Types of 2HDMs, 
then the contribution is the same, giving 
\begin{eqnarray}
\frac{\Gamma^{\rm MFV}_{\gamma\gamma}(\lambda_{h H^\pm H^\pm},M_{H^\pm})}{\Gamma^{\rm 2HDM}_{\gamma\gamma}(\lambda_{h H^\pm H^\pm},M_{H^\pm})} 
  &=& 1 ~, \\
\frac{\Gamma^{\rm MFV}_{Z\gamma}(\lambda_{h H^\pm H^\pm},M_{H^\pm})}{\Gamma^{\rm 2HDM}_{Z\gamma}(\lambda_{h H^\pm H^\pm},M_{H^\pm})} 
  &=& 1 \, .
\label{eq:mfvover2hdmgamgam}
\end{eqnarray}
We must emphasize that the ratios, Eq.~(\ref{eq:mfvover2hdmgamgam}),
are equal only when the charged Higgs coupling and mass are taken to be the same.
As we will see, the \emph{range} of the parameter space 
($\lambda_{h H^\pm H^\pm}$, $M_{H^\pm}$) that is allowed by 
experimental constraints may be considerably wider in the 2HDM Type MFV 
model versus the 2HDM Types I-IV, thus permitting larger effects on these
rates.

Finally, Higgs partial widths into fermions are Type-dependent,
given by
\begin{eqnarray}
\frac{\Gamma^{\rm MFV}_{b\bar{b}}}{\Gamma^{\rm 2HDM}_{b\bar{b}}} &=& 
  \left( \frac{1 - \eps_d / \tan\alpha}{1 + \eps_d \tan\beta} \right)^2
  \frac{\sin^2\alpha}{\cos^2\beta}\frac{1}{(\xi_d^{\rm 2HDM})^2} ~, \\
\frac{\Gamma^{\rm MFV}_{\tau^+\tau^-}}{\Gamma^{\rm 2HDM}_{\tau^+\tau^-}} &=& 
  \left( \frac{1 - \eps_\ell / \tan\alpha}{1 + \eps_\ell \tan\beta} \right)^2
  \frac{\sin^2\alpha}{\cos^2\beta}\frac{1}{(\xi_\ell^{\rm 2HDM})^2} \, . 
\end{eqnarray}
The 2HDM ratios $\xi^{\rm 2HDM}_{d,\ell}$ can be obtained by applying
Eq.~(\ref{eq:2hdmcouplinglimits}) to Eq.~(\ref{eq:reduced_couplings}) 
for the couplings of the light Higgs $h$:
\begin{eqnarray*}
\xi^{\rm 2HDM}_d = \xi^{\rm 2HDM}_\ell = \cos\alpha/\sin\beta & \quad & 
  \mbox{(Type I)} \\
\xi^{\rm 2HDM}_d = \xi^{\rm 2HDM}_\ell = -\sin\alpha/\cos\beta & \quad & 
  \mbox{(Type II)} \\
\xi^{\rm 2HDM}_d = \cos\alpha/\sin\beta, \; 
\xi^{\rm 2HDM}_\ell = -\sin\alpha/\cos\beta & \quad & 
  \mbox{(Type III)} \\
\xi^{\rm 2HDM}_d = -\sin\alpha/\cos\beta, \; 
\xi^{\rm 2HDM}_\ell = \cos\alpha/\sin\beta  & \quad &
  \mbox{(Type IV)} \, . 
\end{eqnarray*}
Here we see that the Type MFV interpolates among all other 
Types of 2HDMs following the limits in Eq.~(\ref{eq:2hdmcouplinglimits}). 
There are, however, several fascinating parameter ranges
that \emph{are not} reached in any of these models.
These include:
\begin{eqnarray}
\eps_{d}    \; \simeq & \tan\alpha \; : \quad &
    \Gamma_{b\bar{b}}^{\rm MFV} \ll \Gamma_{b\bar{b}}^{\rm 2HDM} 
    \nonumber \\
\eps_{\ell} \; \simeq & \tan\alpha \; : \quad &
    \Gamma_{\tau^+\tau^-}^{\rm MFV} \ll \Gamma_{\tau^+\tau^-}^{\rm 2HDM} 
    \nonumber \\
\eps_{d} \; \simeq & -1/\tan\beta \; : \quad &
    \Gamma_{b\bar{b}}^{\rm MFV} \gg \Gamma_{b\bar{b}}^{\rm 2HDM} 
    \nonumber \\
\eps_{\ell} \; \simeq & -1/\tan\beta \; : \quad &
    \Gamma_{\tau^+\tau^-}^{\rm MFV} \gg \Gamma_{\tau^+\tau^-}^{\rm 2HDM}  
\end{eqnarray}
Note that close to the limit $\eps_{d,\ell} \to -1/\tan\beta$, 
the Yukawa couplings can become non-perturbatively large.

Given that $\Gamma_{b\bar{b}}$ is the dominant part of the total 
width of the lightest Higgs boson, these effects can have 
dramatic consequences on all of the resulting 
Higgs branching ratios.

\subsection{Differences between the 2HDM Type MFV and Types I-IV}

In summary, there are two central differences between 
2HDM Type MFV versus Types I-IV: 
\begin{itemize}
\item The width $\Gamma^{\rm MFV}_{b\bar{b}}$ can be completely different
from $\Gamma^{\rm 2HDM}_{b\bar{b}}$ (and $\Gamma^{\rm SM}_{b\bar{b}}$).
\item The width $\Gamma^{\rm MFV}_{\tau^+\tau^-}$ can be completely different
from $\Gamma^{\rm 2HDM}_{\tau^+\tau^-}$ (and $\Gamma^{\rm SM}_{\tau^+\tau^-}$).
\end{itemize}
Since the total width of the light Higgs boson, $\Gamma_{\rm tot}$, 
is dominated by $\Gamma_{b\bar{b}}$ for $\mh = 125$~GeV, 
this leads to the other important difference:
\begin{itemize}
\item The total width $\Gamma^{\rm MFV}_{\rm tot}$ can be significantly
smaller or larger than $\Gamma^{\rm 2HDM}_{\rm tot}$ 
(and $\Gamma^{\rm SM}_{\rm tot}$).
\end{itemize}

The total width affects all of the branching fractions in a 
correlated way.  To a very good approximation, 
$\Gamma_{\tau^+\tau^-} \ll \Gamma_{b\bar{b}}$ remains true for any 
2HDM modification that can fit the ATLAS and CMS data.
This is simply because the convolution of the production and decay 
of the Higgs boson to $\tau^+\tau^-$ must be less than roughly 
the standard model rate.  Hence, 
the total width ratio can be estimated as
\begin{equation}
\frac{\Gamma^{\rm MFV}_{\rm tot}}{\Gamma^{\rm 2HDM}_{\rm tot}} = 
1 + \left[ \left( 
           \frac{-\sin\alpha + \eps_d \cos\alpha}{\cos\beta + \eps_d \sin\beta} 
           \right)^2 
           - \left( \xi^{\rm 2HDM}_d \right)^2 \right]
    \frac{\Gamma^{\rm SM}_{b\bar{b}}}{\Gamma^{\rm 2HDM}_{\rm tot}} .
\end{equation}

\section{One Light Higgs Boson} \label{sec:h}

We now confront the 2HDM Type MFV with the available data on standard model 
Higgs searches.  We first consider a scenario with one light scalar 
Higgs boson at $M_h \simeq 125$~GeV\@.
The second scalar boson $H$ is assumed to be heavier than 
$M_H \gtrsim 150$~GeV, such that it does not directly influence the 
interpretation of the data. A scenario with two light scalar bosons 
is discussed in Sec.~\ref{sec:hH} below.

\begin{figure*}[tb]
\centering
\includegraphics[width=0.98\textwidth]{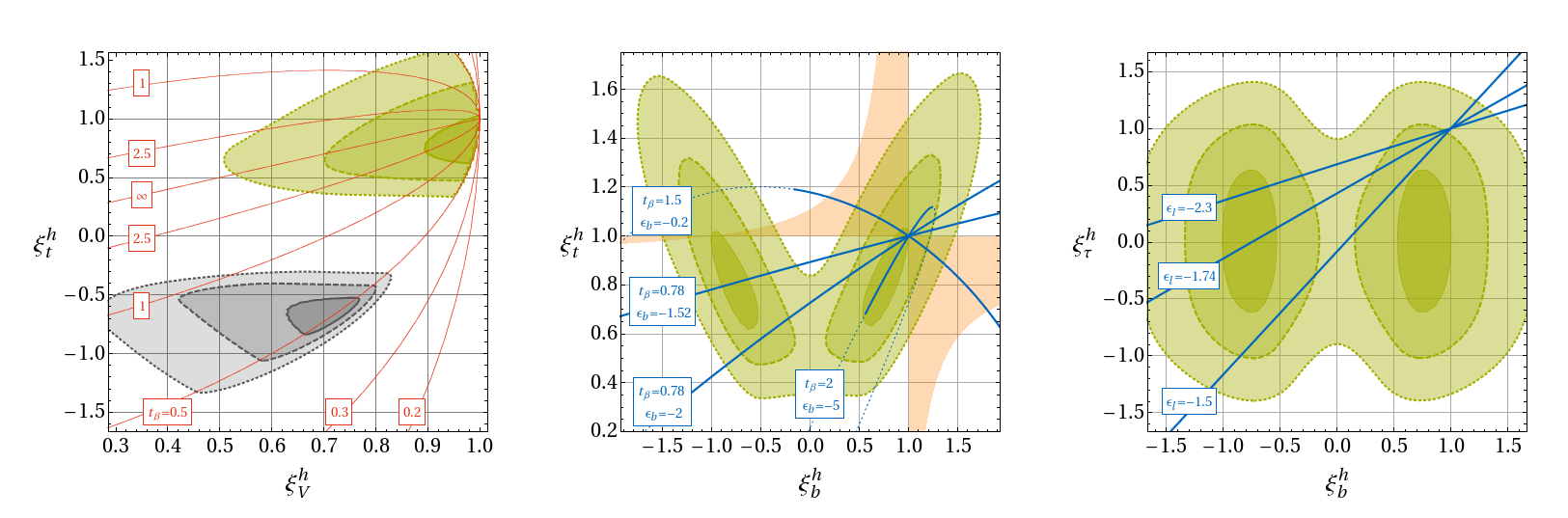}
\caption{Best fit regions in the $\xi_V^h$ -- $\xi_t^h$ (left), $\xi_b^h$ -- $\xi_t^h$ (middle), and $\xi_b^h$ -- $\xi_\tau^h$ (right) planes in a $\chi^2$ fit of the data to one Higgs boson at 125~GeV\@. The light green, green and dark green regions correspond to the $\Delta \chi^2 =$1, 4, and 9 regions. The red labeled contours in the left plot show constant values of $\tan\beta$. In the middle plot the region shaded in orange shows the parameter space that is accessible in a 2HDM Type II by varying $\xi_V^h$ within its $1\sigma$ range. The blue solid curves in the middle and right plot exemplarily show regions of parameter space that can be reached in the MFV 2HDM by varying $\xi_V^h$ within the $1\sigma$ range, while keeping the other parameters fixed to the indicated values (dotted lines correspond to $\xi_V^h$ outside the $1\sigma$ range). For the blue curves in the right plot we fix $\tan\beta = 0.78$ and $\epsilon_b = -1.52$ to the best fit values. The gray region in the left plot with $\xi_t^h < 0$ corresponds to a second minimum in the $\chi^2$, that is however excluded by searches for the heavy scalar $H$ (see text).}
\label{fig:fit_h}
\end{figure*}

\begin{figure}[tb]
\centering
\includegraphics[width=0.48\textwidth]{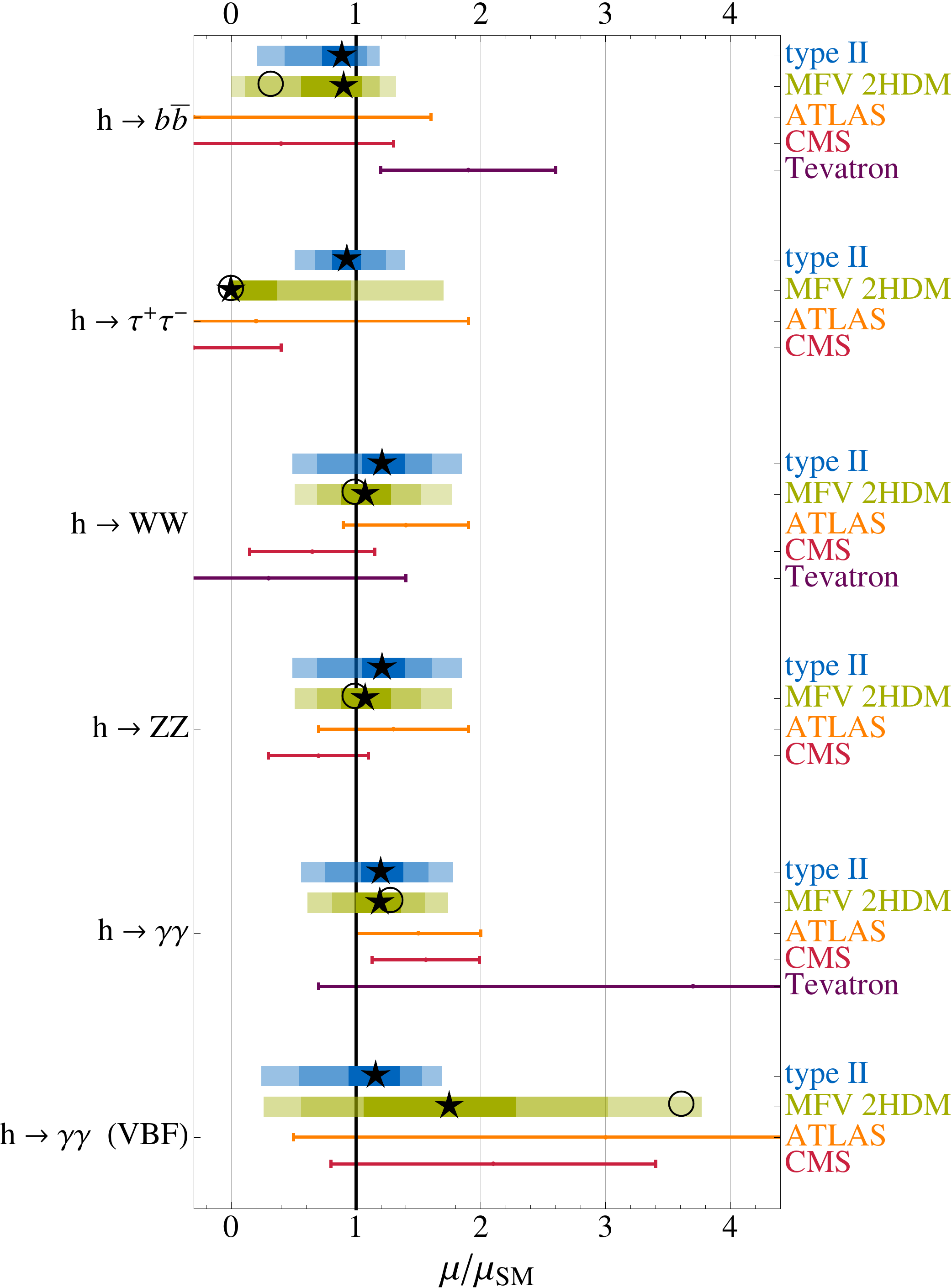}
\caption{Results for various Higgs rates normalized to the SM rates from a fit of the data to the 2HDM Type MFV with one light scalar boson at 125~GeV\@. For comparison results from an analogous fit in the 2HDM Type II and the experimental $1\sigma$ ranges are also shown. The black stars correspond to the best fit values. The circles indicate an example scenario with strongly enhanced VBF $h \to \gamma \gamma$ signal.}
\label{fig:fit_results_h}
\end{figure}

\subsection{Fit to the Data} \label{sec:fit_h}

To determine the preferred values of the couplings of the 2HDM Type MFV given the current data, we perform a simple $\chi^2$ fit of $\xi_V^h$, $\xi_t^h$, $\xi_b^h$, and $\xi_\tau^h$ taking into account SM Higgs searches at LHC~\cite{:2012gk,:2012gu,ATLAS-CONF-2012-127} and at Tevatron~\cite{:2012cn}. 
Similar fits have also been performed in~\cite{Batell:2011pz,Carmi:2012yp,Azatov:2012bz,Espinosa:2012ir,Giardino:2012ww,Low:2012rj,Klute:2012pu,Azatov:2012wq,Giardino:2012dp,Buckley:2012em,Ellis:2012hz,Espinosa:2012im,Plehn:2012iz}.
We consider:
\begin{itemize}
\item[(a)] searches for $h \to b\bar{b}$ produced in association with a gauge boson from ATLAS~\cite{ATLAS-CONF-2012-015}, CMS~\cite{CMS-PAS-HIG-12-019}, and Tevatron~\cite{:2012cn},
\item[(b)] searches for $h \to b\bar{b}$ produced in association with top quarks by CMS~\cite{CMS-PAS-HIG-12-025} and ATLAS~\cite{ATLAS-CONF-2012-135},
\item[(c)] the inclusive $h \to \tau^+\tau^-$ searches at ATLAS~\cite{ATLAS-CONF-2012-014} and CMS~\cite{CMS-PAS-HIG-12-018},
\item[(d)] the inclusive $h \to WW$ searches from ATLAS~\cite{ATLAS-CONF-2012-098}, CMS~\cite{CMS-PAS-HIG-12-017}, and Tevatron~\cite{:2012cn},
\item[(e)] the results on $h \to WW$ produced in VBF by CMS~\cite{Chatrchyan:2012ty} and ATLAS~\cite{ATLAS-CONF-2012-127},
\item[(f)] the inclusive $h \to ZZ \to 4\ell$ searches from ATLAS~\cite{ATLAS-CONF-2012-092} and CMS~\cite{CMS-PAS-HIG-12-016},
\item[(g)] the inclusive $h \to \gamma\gamma$ searches from ATLAS~\cite{ATLAS-CONF-2012-091}, CMS~\cite{CMS-PAS-HIG-12-015}, and Tevatron~\cite{:2012cn},
and finally 
\item[(h)] the results on $h \to \gamma\gamma$ produced in VBF by CMS~\cite{CMS-PAS-HIG-12-015} and ATLAS~\cite{ATLAS-CONF-2012-127}.
\end{itemize}

In a 2HDM the couplings of the Higgs boson to gauge bosons is constrained to be $\xi_V^{h} \le 1$. Furthermore certain regions of the parameter space are only accessible for very small values of $\tan\beta$ that lead to non-perturbative $\xi_u^A$ and $\xi_u^-$ couplings (see Eq.~(\ref{eq:reduced_couplings})).
We therefore perform the fit imposing the constraints $\xi_V^h < 1$ and $\tan\beta > 0.5$.\footnote{We checked that none of our conclusion changes by allowing for even lower values of $\tan\beta \gtrsim 0.3$.} The resulting best fit regions of parameter space are shown in Fig.~\ref{fig:fit_h} in the $\xi_V^h$ -- $\xi_t^h$ plane (left plot), the $\xi_b^h$ -- $\xi_t^h$ plane (center plot), and the $\xi_b^h$ -- $\xi_\tau^h$ plane (right plot). 
The dark green, green and light green regions correspond to $\Delta \chi^2 =$1, 4, and 9, respectively, and we will refer to them as 1, 2 and 3$\sigma$ regions.
Throughout each plot the other couplings are chosen to minimize the total $\chi^2$.

In the best fit region, the $\xi_V^h$ coupling is to a good approximation SM-like while the $\xi_t^h$ coupling is reduced but still positive. The reduced $\xi_t^h$ gives a slight enhancement of the partial width of $h \to \gamma\gamma$. Simultaneously, it also reduces the gluon fusion production cross section. Therefore, in order to obtain an enhanced inclusive $h \to \gamma\gamma$ rate as hinted by the data, a reduction of the total width is also required. This is achieved by reducing the $\xi_b^h$ coupling as shown in the center plot of Fig.~\ref{fig:fit_h}. Both signs of the $\xi_b^h$ coupling are allowed and give essentially equivalent results for the light Higgs boson. Finally as shown in the right plot of Fig.~\ref{fig:fit_h}, a reduced $\xi_\tau^h$ coupling is preferred because it leads to a strongly reduced $h \to \tau^+\tau^-$ signal, as hinted by CMS data~\cite{CMS-PAS-HIG-12-018}.

The best fit values that we find read
\begin{eqnarray}
\xi_V^h = 0.99 ~&,&~~ \xi_t^h = 0.79~, \nonumber \\
\xi_b^h = \pm 0.73 ~&,&~~ \xi_\tau^h = 0~.
\end{eqnarray}
We remark that there exists another minimum in the $\chi^2$ in a region where the $\xi_t^h$ coupling has the opposite sign compared with the SM\@. This region is shown in gray in the left plot of Fig.~\ref{fig:fit_h}. In this region one could expect a considerable enhancement of the $h \to \gamma\gamma$ partial width as W and top loops now interfere constructively.  However, since the Higgs coupling with $W$s is generically suppressed, the enhancement of the di-photon width is small or even absent. Furthermore, as is evident from the left plot of Fig.~\ref{fig:fit_h}, the opposite sign solution for $\xi_t^h$ requires $\tan\beta$ as small as possible. The requirement $\tan\beta \gtrsim 0.5$ then necessarily implies that the $\xi_V^h$ coupling is considerably reduced in magnitude compared to the SM values. 
For $\tan\beta < 1$ the sum rule, Eq.~(\ref{eq:sumrule_u}), implies that the heavy scalar bosons $H$ has an enhanced gluon fusion production cross section. Furthermore, due to the sum rule, Eq.~(\ref{eq:sumrule_V}), both scalar bosons couple in a non-negligible way to vector bosons. Correspondingly, $H$ has a large rate in $H \to WW/ZZ$. Therefore SM Higgs searches rule out the region with $\xi_t^h < 0$ for masses up to $M_H < 600$~GeV, the present limit of the experimental search sensitivities. 
For $M_H > 600$~GeV, the model is already in the decoupling limit, where we expects all light Higgs boson couplings to be SM-like.  We will not consider this region further in this paper. 

Fig.~\ref{fig:fit_results_h} shows the rates of the light Higgs boson in the best fit region with $\xi_t^h > 0$ in comparison to the experimental data. The dark green, green, and light green bands correspond to the 1, 2, and 3$\sigma$ regions of the fit. The black stars mark the best fit values.
The rates obtained for the best fit values follow the data closely: the $h \to \tau^+\tau^-$ rate is reduced to near zero; the largest enhancement is realized for $h \to \gamma\gamma$ in VBF, followed by the inclusive $h \to \gamma\gamma$; the $h \to WW$ and $h \to ZZ$ rates are only slightly enhanced; the $h \to b\bar{b}$ rate is reduced compared to the SM, which is in slight tension with the Tevatron data. Ignoring the Tevatron $h \to b\bar{b}$ data in the fit would allow even larger $h \to \gamma\gamma$ rates, by reducing $h \to b\bar{b}$ further.

\subsection{Comparison with the 2HDM Type II}

In Fig.~\ref{fig:fit_results_h} we also compare the best fit values for the rates of the light Higgs in the 2HDM Type MFV with the corresponding rates in a 2HDM Type II\@. The most important differences are in the inclusive $h \to \tau^+\tau^-$ rate and the VBF $h \to \gamma\gamma$ rate. 
The $\tau^+\tau^-$ rate cannot be reduced to zero in the 2HDM of Type II
in contrast to the 2HDM Type MFV\@.
In a Type II model $\xi_\tau^h = \xi_b^h$ and therefore a strongly suppressed $h \to \tau^+\tau^-$ rate implies a strongly reduced $h\to b\bar{b}$ width. This in turn would lead to a drastic enhancement of all other branching ratios and correspondingly to $h \to ZZ$ and $h \to WW$ rates far above what is allowed by current data.
In the 2HDM Type MFV, instead, $\xi_\tau^h$ and $\xi_b^h$ are independent from each other and $\xi_\tau^h$ is unconstrained by rates other than $h \to \tau^+\tau^-$.

Concerning the VBF $h \to \gamma\gamma$ rate, we observe that in the MFV model huge enhancements are possible.  In Fig.~\ref{fig:fit_results_h} we show an example scenario (open circles) leading to an enhancement by a factor of $\sim 3.5$ that can be realized by the couplings
\begin{eqnarray}
\xi_V^h = 0.97 ~&,&~~ \xi_t^h = 0.49~, \nonumber \\
\xi_b^h = 0.33 ~&,&~~ \xi_\tau^h = 0~.
\end{eqnarray}
By contrast, in the Type II model, we find the VBF $h \to \gamma\gamma$ rate is bounded by approximately $1.7$ at the 3$\sigma$ level.
This difference can be traced back to the strong correlation between $\xi_t^h$ and $\xi_b^h$ in the Type II model. An enhancement of the VBF $h \to \gamma\gamma$ rate by factors of a few is only possible if the $h \to b\bar{b}$ width is reduced considerably, by reducing $\xi_b^h$. This not only enhances the  $h \to \gamma\gamma$ branching ratio but also all other branching ratios. To keep the inclusive $h \to WW$, $h \to ZZ$, and $h \to \gamma\gamma$ rates at a level compatible with experimental data, the gluon fusion production cross section has to be reduced by reducing the $\xi_t^h$ coupling. In contrast to the MFV model where $\xi_t^h$ and $\xi_b^h$ are independent, this is not possible in a Type II model, where a strongly modified $\xi_b^h$ coupling implies $\xi_t^h \simeq 1$ and vice versa.

We remark that an enhancement of the VBF $h \to \gamma\gamma$ rate by a factor of few in the MFV model also implies a similar enhancement of the VBF $h \to WW$ rate. This is in tension with the CMS and ATLAS analyses of $h \to WW \to \ell\nu\ell\nu$ that do not see any excess above background in the $h$ + 2 jets sample that is dominated by VBF production. Note that these results are included in our fit, but given the considerable uncertainties, they do not influence the fit by much. Updated results for the VBF $h \to WW$ rate from ATLAS and CMS can either rule out or support the possibility of an strongly enhanced VBF $h \to WW$ rate in the 2HDM Type MFV\@. 
We also note that a strongly enhanced $Wh \to WWW$ rate is predicted in the regions of parameter space with a strongly enhanced VBF $h \to \gamma\gamma$ rate. However, the corresponding CMS and ATLAS searches~\cite{ATLAS-CONF-2012-078,CMS-PAS-HIG-11-034} currently give only very mild constraints on this channel.

\subsection{Generic Predictions for the Light Higgs}

\begin{table}[tbh]
\renewcommand{\arraystretch}{1.7}
\renewcommand{\tabcolsep}{6pt}
\small
\begin{center}
\begin{tabular}{ccc}
\hline
 & $\mu_{h\to b\bar{b}}^{\rm assoc.}$ & $\mu_{h\to VV}^{\rm incl.}$ \\ \hline\hline
$\mu_{h\to\gamma\gamma}^{\rm incl.} > 1.2$ ~$\Rightarrow$ & $<1$ & $>0.9$ \\
$\mu_{h\to\gamma\gamma}^{\rm incl.} > 1.5$ ~$\Rightarrow$ & $<0.8$ & $>1.2$ \\
\hline
\\ \\
\hline
 & $\mu_{h\to b\bar{b}}^{\rm assoc.}$ & $\sigma_H^{\rm incl.}/{\rm SM}$ \\ \hline\hline
$\mu_{h\to\gamma\gamma}^{\rm VBF} > 2$ ~$\Rightarrow$ & $<0.9$ & $>0.3$ \\
$\mu_{h\to\gamma\gamma}^{\rm VBF} > 3$ ~$\Rightarrow$ & $<0.4$ & $>0.7$ \\
\hline
\end{tabular}
\end{center}
\caption{Correlations between several Higgs rates and cross sections in the 2HDM Type MFV with one light Higgs boson at 125~GeV\@.}
\label{tab:predictions} 
\end{table}

In the 2HDM Type MFV, even though the couplings of the light Higgs to the different Types of fermions can be modified independently, there exists correlations among the rates of some of the currently investigated Higgs search channels.
In Table~\ref{tab:predictions} we summarize generic predictions for the light Higgs that we found.

In particular, we find that an enhancement of the inclusive $h \to \gamma\gamma$ rate implies 
\begin{itemize}
\item[(i)] an upper bound on the $h\to b\bar{b}$ rate where the Higgs is produced in association with a vector boson, and
\item[(ii)] a lower bound on the inclusive $h \to WW$ and $h \to ZZ$ rates.
\end{itemize}
Similarly, also strong enhancements of the VBF $h \to \gamma\gamma$ rate imply stringent upper bounds on the $h\to b\bar{b}$ rate. 
This is clearly shown in Fig.~\ref{fig:fit_results_h} where the circles correspond to an example scenario with the VBF $h \to \gamma\gamma$ rate enhanced by a factor of $\sim 3.5$.
As a very strong enhancement of the VBF $h \to \gamma\gamma$ rate is only viable for a reduced $\xi_t^h$ coupling, the sum rule, Eq.~(\ref{eq:sumrule_u}), implies a lower bound on the production cross section of the heavy scalar $H$, and correspondingly good prospects for $H$ searches at the LHC\@. Note that the bounds presented in Table~\ref{tab:predictions} do not change appreciably if we restrict to scenarios fitting the Higgs data at the 1, 2 or 3$\sigma$ level. Correspondingly, these bounds are robust.

\subsection{The Quasi-Decoupling Limit}

We analyze the extent to which the identified values of the Higgs couplings can be realized concretely in the 2HDM Type MFV\@. In the best fit region with $\xi_t^h$ positive, the coupling of the Higgs with gauge bosons is approximately SM-like. We can express this ``quasi-decoupling limit'' with the relation $\alpha=\beta-\pi/2+x$ where $x$ is a small expansion parameter. The couplings of the Higgs bosons with gauge bosons and fermions are then given by
\begin{eqnarray}
 \xi_V^h &\simeq& 1 - \frac{x^2}{2} ~,\\
 \xi_V^H &\simeq& x ~,\\
  \xi_u^h &\simeq& \left( 1 - \frac{x^2}{2} \right) + x\xi_u^A ~,  \\
 \xi_u^H &\simeq& - \xi_u^A \left( 1 - \frac{x^2}{2} \right) + x ~,  \\\label{eq:39}
 \xi_{d,\ell}^h &\simeq& \left( 1 - \frac{x^2}{2} \right) - x \xi_{d,\ell}^A ~, \\\label{eq:40}
 \xi_{d,\ell}^H &\simeq& \left( 1 - \frac{x^2}{2} \right) \xi_{d,\ell}^A + x ~,
\end{eqnarray}
with the couplings $\xi_f^A$ given in Eq.~(\ref{eq:reduced_couplings}).
We see that even if the coupling of $h$ to weak gauge bosons is SM-like within a few percent, there can be substantial modifications to the remaining couplings (this has been also pointed out recently in~\cite{Alves:2012ez} in the context of the 2HDM Type I\@.). In particular we observe that couplings of the light Higgs with quarks and leptons can be strongly modified if the couplings of the CP-odd Higgs boson with quarks and leptons are sizable.
In this situation the couplings of the heavy CP-even Higgs boson $H$ are to a good approximation the same as the CP-odd Higgs boson $A$.\footnote{Conversely, if the couplings to quarks or leptons of $A$ are small, then the corresponding couplings of $H$ can be modified, with the couplings of $h$ remaining SM-like.}
We again stress that in the MFV framework, the couplings of $h$ to up-type quarks, down-type quarks, and leptons can be modified independently, whereas in 2HDM Types I-IV they are strongly correlated. For example, in the 2HDM Type II, an enhancement of $\xi_d^A = \xi_\tau^A = \tan\beta$ is always accompanied by a reduction of $\xi_u^A = 1/\tan\beta$.

It is impressive that, in contrast to 2HDM Types I-IV, the 2HDM Type MFV can exactly reproduce the best fit values for the Higgs fermion couplings while retaining essentially SM-like couplings of the Higgs to gauge bosons.
This is illustrated in the middle and right plot of Fig.~\ref{fig:fit_h}. 
The additional parameter $\epsilon_b$ in the MFV framework allows us to modify $\xi_t^h$ and $\xi_b^h$ completely independently. Example choices for $\tan\beta$ and $\epsilon_b$ that cover the whole parameter space of top and bottom couplings, and in particular to reach the best fit values, are indicated with the blue solid lines in the middle plot.
Similarly, the right plot of Fig.~\ref{fig:fit_h} shows how the independent parameter $\epsilon_\ell$ can be used to obtain a highly suppressed $\xi_\tau^h$ coupling while fixing $\tan\beta$ and $\epsilon_b$ such that $\xi_t^h$ and $\xi_b^h$ correspond to their best fit values.

For completeness we also report our best fit values for the 2HDM parameters $\tan\beta$, $\alpha$, $\epsilon_b$, and $\epsilon_\tau$. We find
\begin{eqnarray}
\tan\beta = 0.78 ~&,&~~ \nonumber \alpha = -1.05 ~, \nonumber \\
\epsilon_b = -1.52 ~ (-8.3) ~&,&~~ \epsilon_\tau = -1.74~,
\end{eqnarray}
where the two numbers for $\epsilon_b$ correspond to the negative (positive) solution for $\xi_b^h$. In the best fit point, $\tan\beta$ has a small value. From the left plot of Fig.~\ref{fig:fit_h} we note however, that also regions of parameter space with large $\tan\beta$ can result in a very good fit of the data.

\begin{figure}[tb]
\centering
\includegraphics[width=0.48\textwidth]{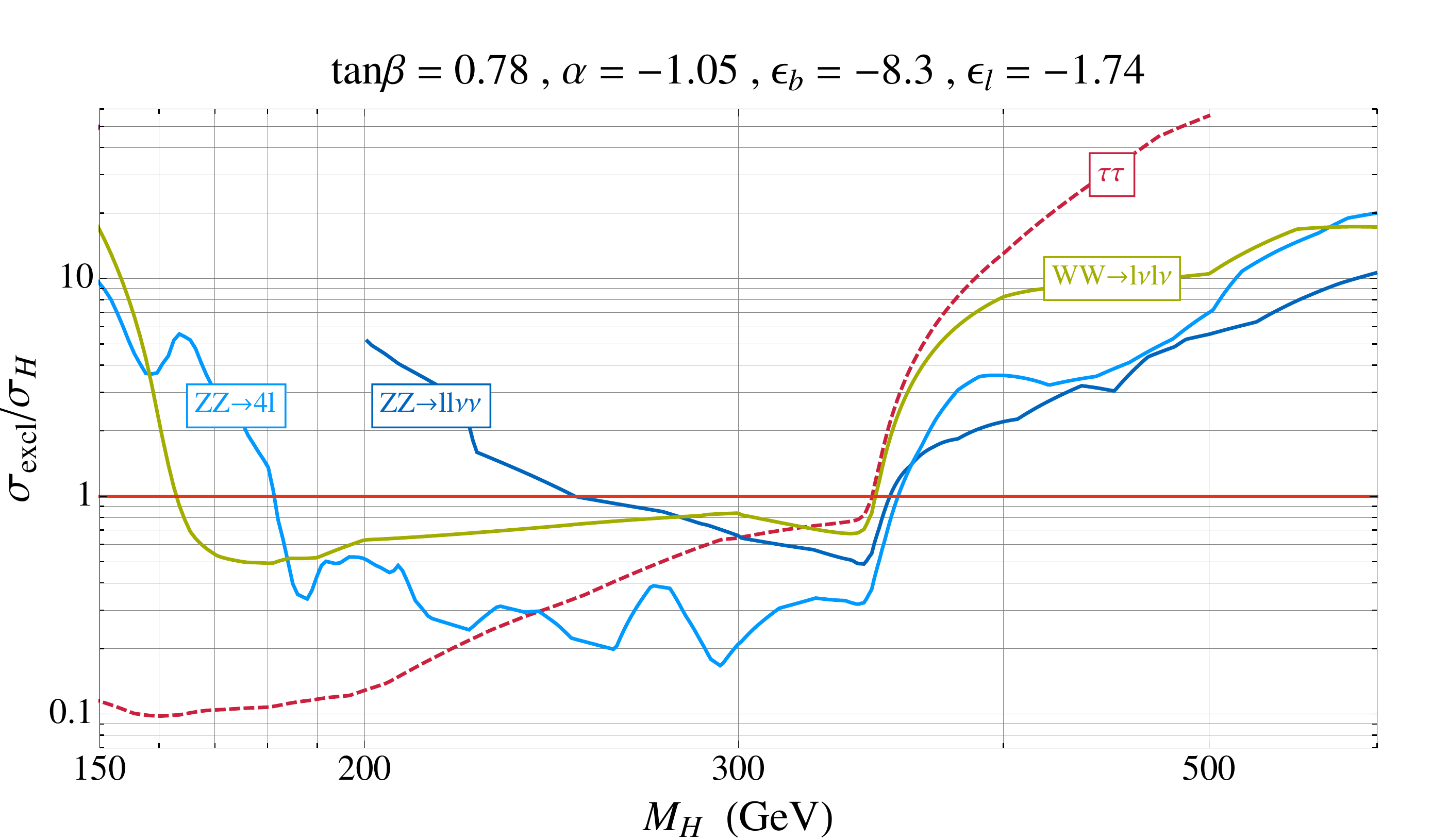} \\[16pt]
\includegraphics[width=0.48\textwidth]{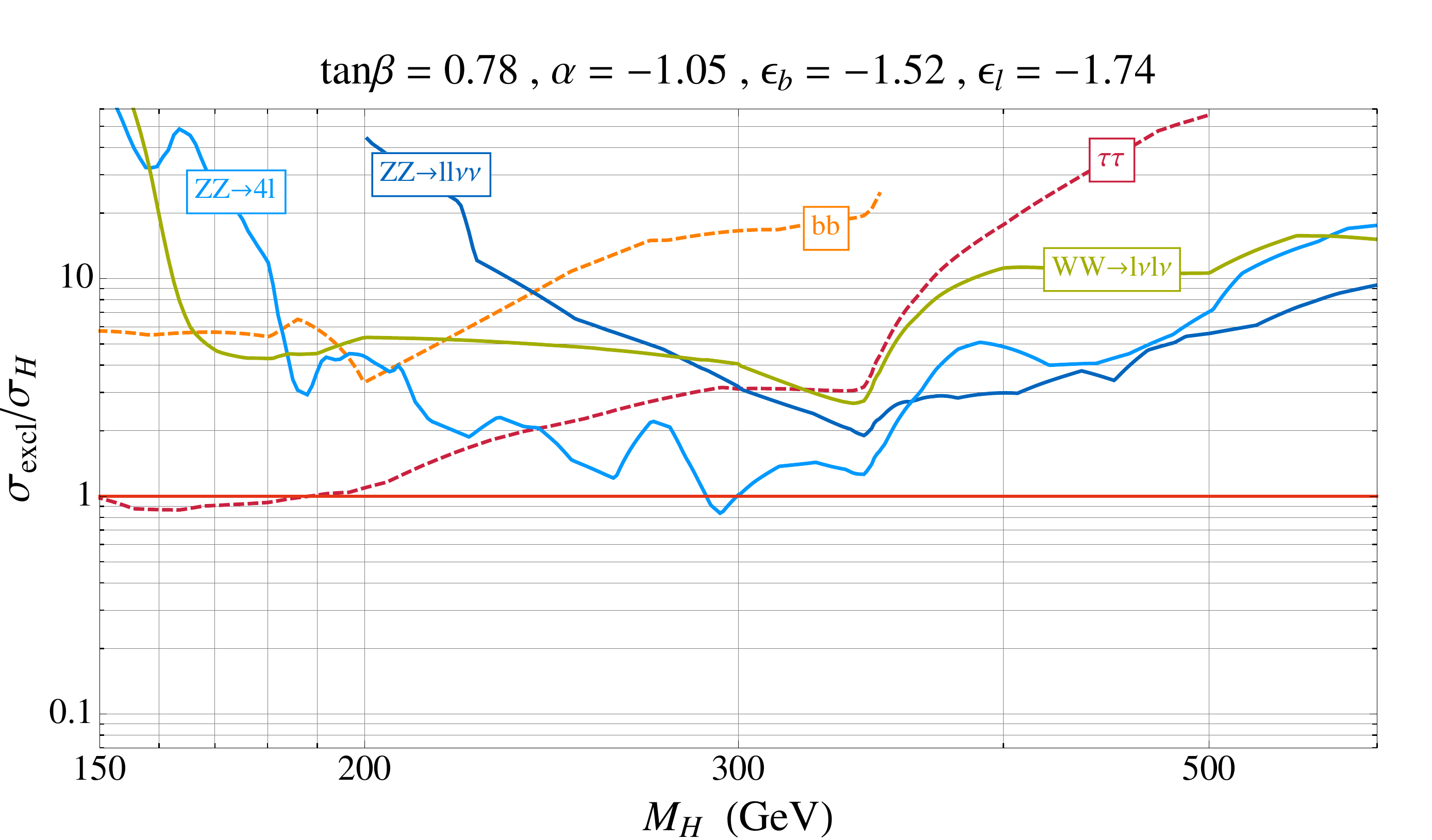} \\[16pt]
\includegraphics[width=0.48\textwidth]{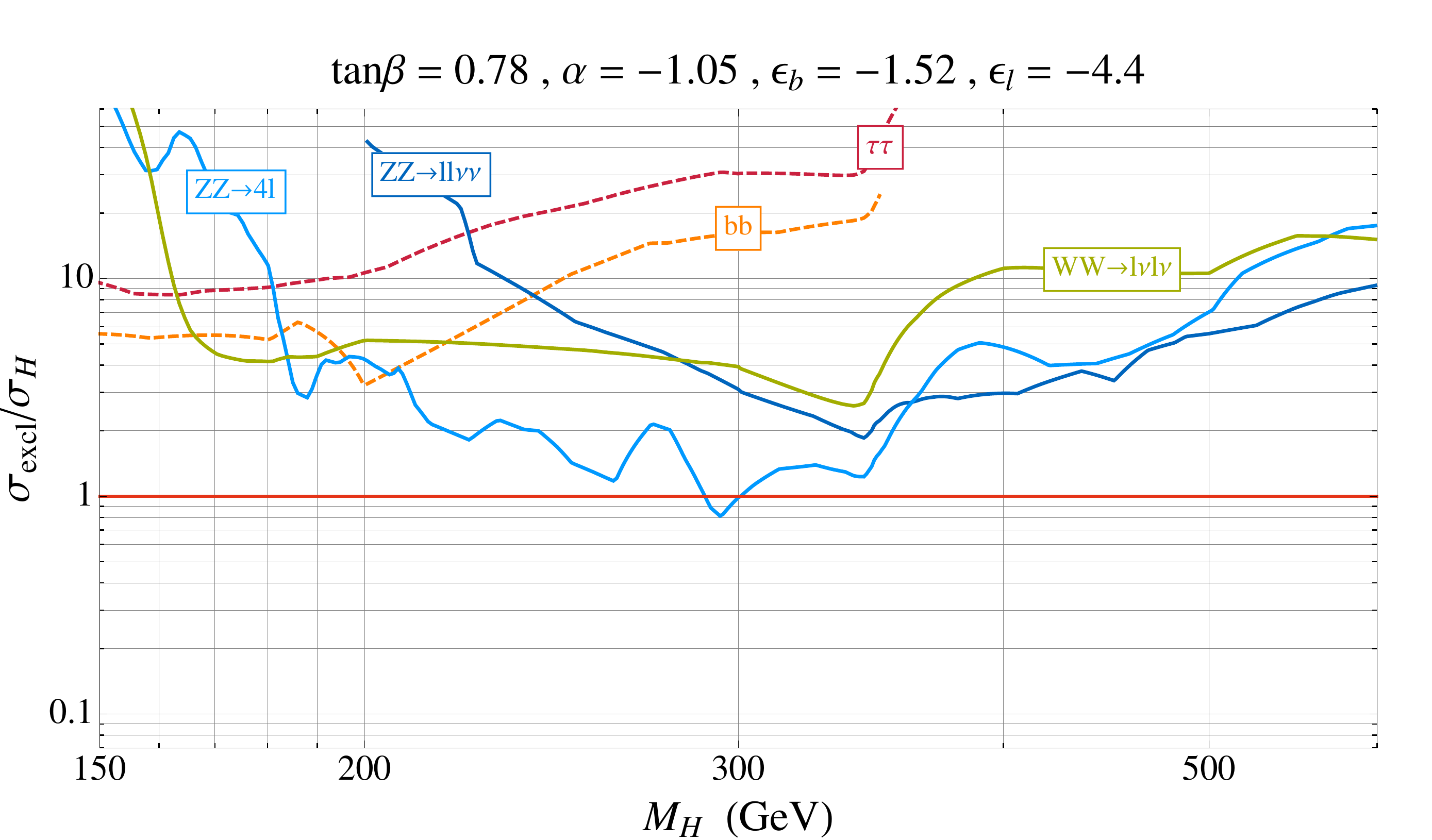}
\caption{Current 95\% C.L. exclusion bounds normalized to the predicted cross sections for the heavy Higgs as function of the heavy Higgs mass $M_H$. Shown are $H \to ZZ \to 4\ell$ (light blue), $H \to ZZ \to \ell \ell \nu\nu$ (dark blue), $H \to WW \to \ell\nu \ell\nu$ (green), $H \to \tau^+\tau^-$ (dark red), and $H \to b\bar{b}$ (orange). The top (center) plot corresponds to the best fit values of the light Higgs couplings with $\xi_b^h >0$ ($\xi_b^h < 0$). In the bottom plot we allow for a larger $\xi_\tau^h$ coupling such that the inclusive $h\to\tau^+\tau^-$ rate is 50\% of the corresponding SM rate.}
\label{fig:predictions_H}
\end{figure}

\subsection{Predictions for the Heavy Scalar \texorpdfstring{$H$}{H}}

Once the couplings of the light Higgs boson are fixed, then the couplings of the heavy Higgs boson $H$ (as well as the CP-odd Higgs boson $A$ and the charged Higgs bosons $H^\pm$) are determined. At the best fit values, we find
\begin{eqnarray} \label{eq:Hcouplings}
\xi_V^H = 0.14 ~&,&~~ \xi_t^H = 1.36~, \nonumber \\
\xi_b^H = -12.4 ~ (-1.78) ~&,&~~ \xi_\tau^H = -7.1~.
\end{eqnarray}
In Fig.~\ref{fig:predictions_H} we show the resulting predictions of various cross sections of the heavier scalar $H$ as a function of its mass for the two solutions for $\xi_b^h$.\footnote{Here we do not consider the possibility of a sizable $H \to hh$ decay rate, as the corresponding coupling is another free parameter of the model.}
The top (center) plot corresponds to the best fit point with $\xi_b^h >0$ ($\xi_b^h <0$). In the bottom plot we allow for a larger $\xi_\tau^h$ coupling such that the inclusive $h\to\tau^+\tau^-$ rate is 50\% of the corresponding SM rate\footnote{As shown in Fig.~\ref{fig:fit_results_h}, $h\to\tau^+\tau^-$ rates around $50\%$ are still resulting in a reasonable good fit of the present Higgs data.}. Plotted are the current exclusion bounds normalized to the predicted signal cross sections. We take into account not only searches for SM-like Higgs in the $H \to ZZ \to 4\ell$~\cite{ATLAS-CONF-2012-092,CMS-PAS-HIG-12-016} (light blue), $H \to ZZ \to \ell \ell \nu\nu$~\cite{:2012va,CMS-PAS-HIG-12-023} (dark blue), and $H \to WW \to \ell\nu \ell\nu$~\cite{Aad:2012sc,CMS-PAS-HIG-12-017} (green) channels, but also searches for MSSM Higgs bosons in the $H \to \tau^+\tau^-$~\cite{Chatrchyan:2012vp,ATLAS-CONF-2012-094}\footnote{Note that the CMS analysis~\cite{Chatrchyan:2012vp} only provides bounds in the $M_A$-$\tan\beta$ plane of a specific MSSM scenario. We translate these bounds into bounds on the signal cross sections and reinterpret them in our scenario assuming constant efficiencies.} (dark red) and $H \to b\bar{b}$~\cite{CMS-PAS-HIG-12-026,CMS-PAS-HIG-12-027} (orange) channels. For every value of $M_H$ we consider the strongest of the individual bounds from ATLAS and CMS.

We find that for the best fit point with $\xi_b^h >0$ (top panel), the heavy scalar is excluded up to $M_H \lesssim 350$~GeV by the current searches.  Once $M_H > 350$~GeV, the heavy scalar has a large branching fraction into $t \bar t$ and is correspondingly only weakly constrained by present data. 
In the $\xi_b^h <0$ case, the coupling of the heavy Higgs to bottom quarks is considerably larger, c.f.~Eq.~(\ref{eq:Hcouplings}). As a result, the branching ratios of $H$ into $\tau^+\tau^-$ and vector bosons are significantly smaller and current searches only start to be sensitive to the region below $M_H \lesssim 350$~GeV\@.
Searches in the di-tau final state are able to just barely exclude the heavy scalar up to $M_H \simeq 200$~GeV\@. Searches for $h \to ZZ \to 4\ell$ start to become sensitive to $H$ in the mass range up to 350~GeV\@. Above 350~GeV, the heavy scalar again decays dominantly into $t \bar t$.
We remark that the reason for the strong constraints from searches in the $\tau^+\tau^-$ final state is the best fit preference to suppress the $h \to \tau^+\tau^-$ rate as much as possible. Indeed a $\xi_\tau^h$ coupling close to zero is only possible with an enhanced coupling of the heavy scalar $H$ to $\tau$'s [see Eq.~(\ref{eq:sumtau})]. If we deviate slightly from the best fit point and allow a somewhat larger $h \to \tau^+\tau^-$ rate, the parameter space for the heavy scalar opens up further. This is shown in the bottom panel of Fig.~\ref{fig:predictions_H} where we fix $\epsilon_\ell$ such that the inclusive $h \to \tau^+\tau^-$ rate is 50\% of the corresponding SM rate. Now the $\xi_\tau^H$ coupling is much smaller and the heavy Higgs is only excluded in a small region around 300~GeV by searches in the $4\ell$ final state. Interestingly, apart from the $h \to \tau^+\tau^-$ rate, all other channels are to an excellent approximation unaffected by changing $\epsilon_\ell$. Note also that in this scenario the current $H \to b\bar b$ constraints are stronger than the $H \to \tau^+\tau^-$ ones, showing that $H \to b\bar b$ MSSM searches at the LHC~\cite{Carena:2012rw} give valuable complementary information in the 2HDM Type MFV.

If the other couplings of the light Higgs are modified from their best fit values, prospects to probe the heavy Higgs typically remain excellent throughout large parts of the parameter space:
\begin{itemize}
\item For a $\xi_t^h$ closer to 1 and correspondingly for larger values of $\tan\beta$, the gluon fusion production cross section for $H$ can be reduced. Still even for large $\tan\beta$ we find that the heavy Higgs signals are typically only a factor of a few above the current exclusion limits. 
\item For larger $\tan\beta$ the coupling of $H$ to the top quark is reduced, implying reduced branching ratios of $H \to t\bar{t}$ for $M_H \gtrsim 350$~GeV and therefore increased sensitivity of $H \to WW/ZZ$ in the large $M_H$ regime.
\item A reduced $\xi_V^h$ coupling leads to a larger $\xi_V^H$ coupling and therefore the $H \to WW$ and $H \to ZZ$ signals get enhanced. The current Higgs searches already probe the corresponding parts of parameter space.   
\item As already discussed, the value of the $\xi_b^h$ coupling strongly influences the branching ratios of $H$. The larger the deviation of $\xi_b^h$ from 1, the larger is the $\xi_b^H$ coupling [see Eqs.~(\ref{eq:39}),(\ref{eq:40}) and the discussion below]. Nonetheless, we find that even the largest deviations of $\xi_b^h$ from 1 generically lead to signals in the $H \to WW/ZZ$ channels that can observed in the near future, as long as $M_H \lesssim 350$~GeV\@.  
\end{itemize}
There also exist also corners of parameter space where the heavy scalar cannot be detected. If the couplings of the light Higgs to the top quark and gauge bosons are to a very high precision SM-like and $\tan\beta$ is large, the couplings of $H$ to the top quark and gauge bosons can be made arbitrarily small. This results in very small production cross sections for $H$ that are not easily detectable.

The phenomenology of the heavy CP-odd Higgs $A$ differs from that of the CP-even $H$, as $A$ does not couple to weak gauge bosons. Thus the only search channels that are currently able to probe $A$ are the $A \to \tau^+\tau^-$ and $A \to b \bar b$ searches. Analogous to the $H$ boson, for $M_A\gtrsim350$~GeV, the $A \to t \bar t$ decay can open up.

\section{Two Light Higgs Bosons} \label{sec:hH}

\begin{figure*}[tb]
\centering 
\includegraphics[width=0.98\textwidth]{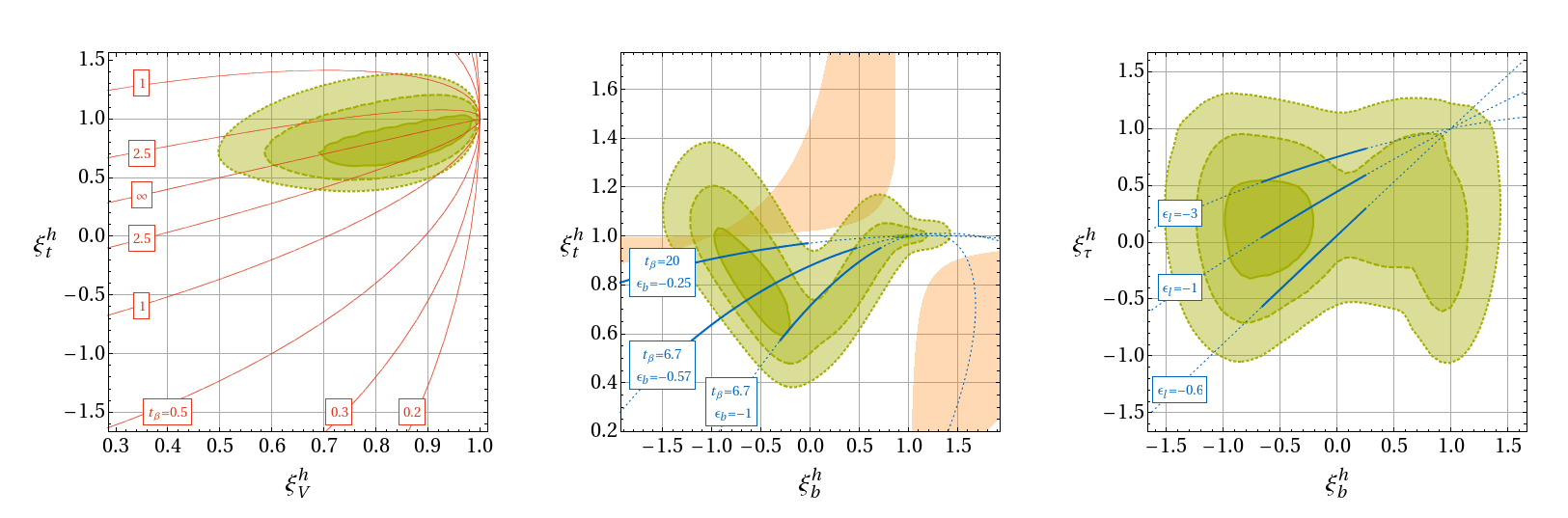}
\caption{Best fit regions in the $\xi_V^h$ -- $\xi_t^h$ (left), $\xi_b^h$ -- $\xi_t^h$ (middle), and $\xi_b^h$ -- $\xi_\tau^h$ (right) planes in a $\chi^2$ fit of the data to two Higgs bosons at 125~GeV and 135~GeV, respectively. The dark green, green and light green regions correspond to the $\Delta \chi^2 =$1, 4, and 9 regions. The red labeled contours in the left plot show constant values of $\tan\beta$. In the middle plot the region shaded in orange shows the parameter space that is accessible in a 2HDM Type II by varying $\xi_V^h$ within its $1\sigma$ range. The blue solid curves in the middle and right plot exemplarily show regions of parameter space that can be reached in the MFV 2HDM by varying $\xi_V^h$ within the $1\sigma$ range, while keeping the other parameters fixed to the indicated values. For the blue curves in the right plot we fix $\tan\beta = 6.7$ and $\epsilon_b = -0.57$ to the best fit values.}
\label{fig:fit_hH}
\end{figure*}

\begin{figure}[tb]
\centering
\includegraphics[width=0.48\textwidth]{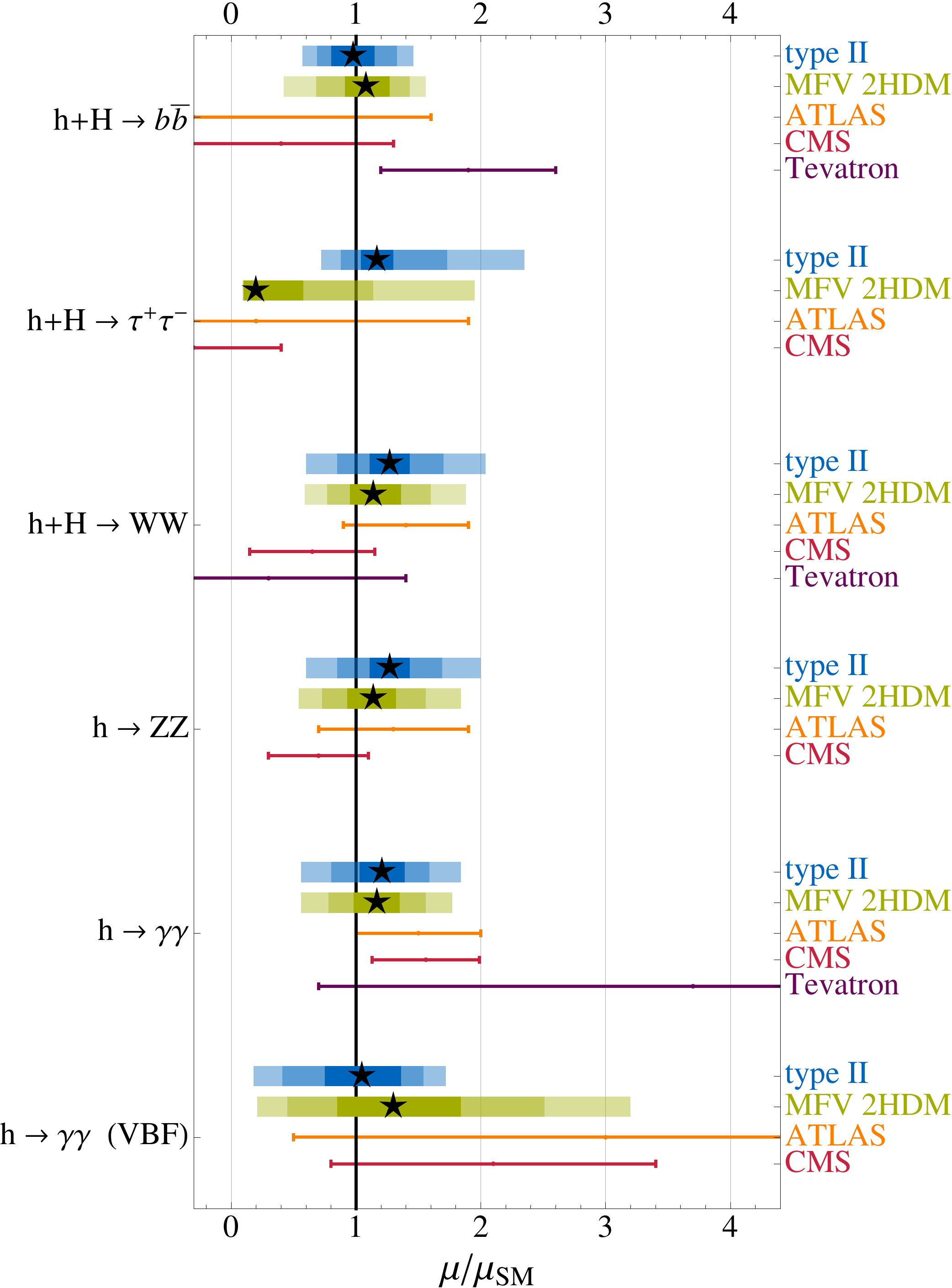}
\caption{Results for various Higgs rates normalized to the SM rates from a fit of the data to the 2HDM Type MFV with two light scalar bosons at 125~GeV and 135~GeV. For comparison, results from an analogous fit in the 2HDM Type II and the experimental 1$\sigma$ ranges are also shown. The black stars correspond to the best fit values.}
\label{fig:fit_results_hH}
\end{figure}

We now investigate a scenario where, in addition to the lightest Higgs boson at 125~GeV, the second CP even Higgs $H$ boson is also light.
For definiteness we fix its mass to $M_H = 135$~GeV\@. We checked that varying the mass of the second Higgs between 130~GeV and 140~GeV does not change the results qualitatively.
For a recent discussion of a similar scenario in the context of the NMSSM see~\cite{Belanger:2012he}.
Scenarios with two quasi-degenerate Higgs bosons at 125~GeV are discussed in~\cite{Gunion:2012gc,Batell:2012mj,Gunion:2012he}.

\subsection{Fit to the Data} \label{sec:fit_hH}

In the Higgs search channels with mass resolution smaller than $M_H - M_h \simeq 10$~GeV, specifically $h \to b\bar{b}$, $h \to \tau^+\tau^-$ and $h \to WW$, the LHC collaborations would effectively be observing the sum over the signals coming from both Higgs bosons $h$ and $H$. For the high resolution channels $h \to \gamma \gamma$ and $h \to ZZ$, we instead consider the data on the signal strength separately for $h$ and $H$.

The resulting best fit regions in the $\xi_V^h$ -- $\xi_t^h$, $\xi_b^h$ -- $\xi_t^h$, and $\xi_b^h$ -- $\xi_\tau^h$ planes are shown in Fig.~\ref{fig:fit_hH}. 
Interestingly enough, now that the effects of the second Higgs boson are directly included in the fit, in the $\xi_V^h$ -- $\xi_t^h$ plane only one solution with $\xi_t^h > 0$ is present. Both couplings are slightly reduced compared to the SM values. 
The two signs of the $\xi_b^h$ coupling are not equivalent anymore. We find a slight preference for a negative $\xi_b^h$ that is below 1 in magnitude. The preferred $\xi_\tau^h$ coupling remaining close to zero.
The best fit point is given by
\begin{eqnarray}
\xi_V^h = 0.85 ~&,&~~ \xi_t^h = 0.77~, \nonumber \\
\xi_b^h = -0.52 ~&,&~~ \xi_\tau^h = 0.16~.
\end{eqnarray}
The resulting couplings of the heavier Higgs are 
\begin{eqnarray}
\xi_V^H = 0.53 ~&,&~~ \xi_t^H = 0.66~, \nonumber \\
\xi_b^H = -2.7 ~&,&~~ \xi_\tau^H = -1.6~.
\end{eqnarray}

The reduced couplings of the heavier Higgs to the top quark and vector bosons lead to suppressed production cross sections of $H$. The enhanced coupling to bottom quarks, however, results in a branching ratio of BR$(H \to b\bar{b}) \simeq 95\%$. Correspondingly, the second Higgs primarily just adds to the $h/H \to b\bar{b}$ signal while its effect in the other search channels is negligible to a first approximation.

The best fit values can be accommodated in the 2HDM Type MFV for appropriate choices of $\epsilon_b$ and $\epsilon_\ell$. Example choices for $\tan\beta$,  $\epsilon_b$, and $\epsilon_\ell$ that allow to cover the whole parameter space of top, bottom and tau couplings, and in particular that reach the best fit values, are indicated with the blue solid lines in the middle and right plot of Fig.~\ref{fig:fit_hH}. In the right plot, $\tan\beta$ and $\epsilon_b$ are fixed such that $\xi_t^h$ and $\xi_b^h$ correspond to their best fit values.
We obtain the best fit values
\begin{eqnarray}
\tan\beta = 6.7 ~&,&~~ \nonumber \alpha = -0.71 ~, \nonumber \\
\epsilon_b = -0.57 ~&,&~~ \epsilon_\tau = -1.0~.
\end{eqnarray}

By contrast, in the 2HDM Types I-IV, the best fit values cannot be accommodated, even if these models can still produce a reasonable good fit of the present LHC Higgs data. The two bands shaded in orange in the middle plot of Fig.~\ref{fig:fit_hH} show the region that can be reached in a 2HDM Type II, if the $\xi_V^h$ coupling is varied in the 1$\sigma$ range around the best fit value. This region does not cover the best fit values for $\xi_b^h$, $\xi_t^h$ and $\xi_\tau^h$.

The Higgs rates resulting from the fit of the two light Higgs boson scenario are compared to the experimental data and to the corresponding fit in the 2HDM Type II in Fig.~\ref{fig:fit_results_hH}.
The best fit points are shown as black stars, while the 1, 2, and 3$\sigma$ ranges are indicated by the green shaded bands.
In the best fit point the $h\to\gamma \gamma$ and $h\to VV$ rates are slightly enhanced due to the suppression of the $\xi_b^h$ coupling that controls the total width of $h$. The dramatic reduction of the $\xi_\tau^h$ coupling leads to a $h\ra \tau^+\tau^-$ rate close to zero. Note however that in contrast to the one Higgs case, the $\tau^+\tau^-$ rate cannot be eliminated completely, because the second scalar $H$ always contributes at some level.
The main difference from the single light Higgs boson case (Sec.~\ref{sec:h}) is that the $h\to b\bar{b}$ rate is slightly \emph{enhanced} compared to the SM prediction, with $h$ and $H$ decays contributing approximately 65\% and 45\% of the SM rate.
At the 3$\sigma$ level, the $b\bar{b}$ rate can even be enhanced by up to a factor of 1.6.
Overall, we find that the fit with the two light Higgs bosons is just as good as the scenario with only one light Higgs boson.

\subsection{Generic Predictions for Higgs Signals}

\begin{table}[tbh]
\renewcommand{\arraystretch}{1.7}
\renewcommand{\tabcolsep}{6pt}
\small
\begin{center}
\begin{tabular}{ccc}
\hline
 & $\mu_{h+H\to b\bar{b}}^{\rm assoc.}$ & $\mu_{h+H\to WW}^{\rm incl.}$  \\ \hline\hline
$\mu_{h\to\gamma\gamma}^{\rm incl.} > 1.2$ ~$\Rightarrow$ & $<1.2$ & $>1.0$ \\
$\mu_{h\to\gamma\gamma}^{\rm incl.} > 1.5$ ~$\Rightarrow$ & $<1.0$ & $>1.3$ \\
\hline
\\ \\
\hline\hline
 & $\mu_{h+H\to b\bar{b}}^{\rm assoc.}$ &  \\ \hline
$\mu_{h\to\gamma\gamma}^{\rm VBF} > 2$ ~$\Rightarrow$ & $< 1.0$ & \\
$\mu_{h\to\gamma\gamma}^{\rm VBF} > 3$ ~$\Rightarrow$ & $< 0.7$ & \\
\hline
\end{tabular}
\end{center}
\caption{Correlations between several Higgs rates in the 2HDM Type MFV with two light Higgs bosons at $M_h = 125$~GeV and $M_H = 135$~GeV.}
\label{tab:predictions_Hh} 
\end{table}

We now discuss generic correlations for the Higgs signals in the 2HDM Type MFV with two light scalar bosons.
As shown in Table~\ref{tab:predictions_Hh}, we find that just like the one light Higgs boson scenario, an enhanced inclusive $h \to \gamma\gamma$ rate implies: 
\begin{itemize}
\item[(i)] an upper bound on the $h+H\to b\bar{b}$ rate where the Higgses are produced in association with a vector boson, and 
\item[(ii)] a lower bound on the inclusive $h+H \to WW$ rate. 
\end{itemize}
Quantitatively however, because of the additional contribution of the second scalar $H$, both the upper bound on the $b\bar{b}$ rate and the lower bound on the $WW$ rate are slightly larger than in the one Higgs case.
Analogously, the upper bounds on the $h+H\to b\bar{b}$ rate that are implied by strong enhancements of the VBF $h \to \gamma\gamma$ rate are weaker compared to the one Higgs case.

Note that in this setup a second peak at the mass of the heavier Higgs is expected in the $H \to \gamma\gamma$ and $H \to ZZ$ channels. However, given the suppressed $\xi_V^H$ and $\xi_t^H$ couplings, the corresponding signal strengths are often only few percent of the SM signals. Therefore, finding evidence for a second light Higgs in these channels is challenging.

\section{Impact of the Charged Higgs Boson}

\begin{figure}[tb]
\centering
\includegraphics[width=0.45\textwidth]{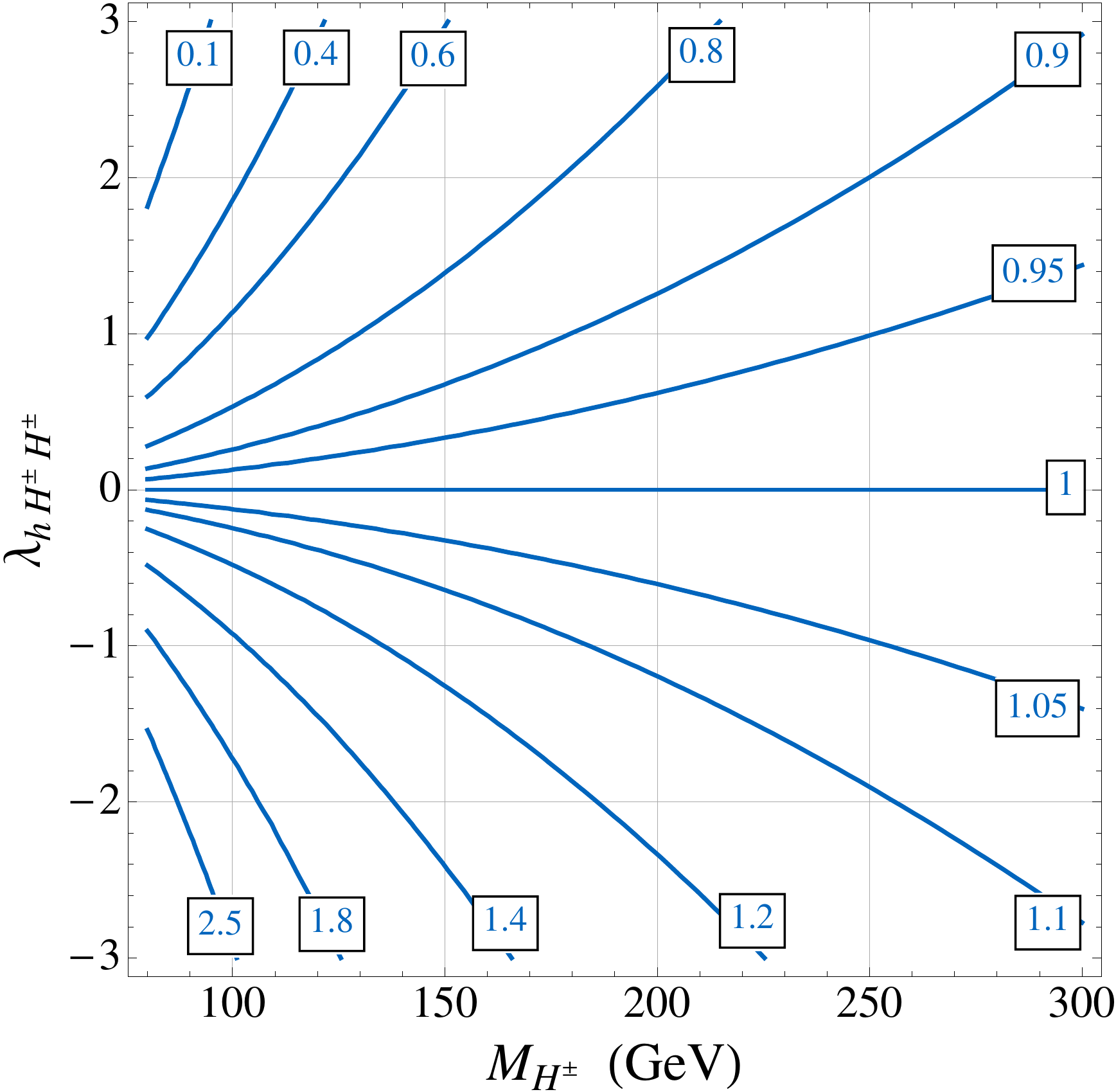}
\caption{Values of the $h \to \gamma\gamma$ rate, normalized to the SM rate in the $M_{H^\pm}$ -- $\lambda_{hH^\pm H^\pm}$ plane. Tree level couplings of the light Higgs to gauge bosons and fermions are assumed to be SM-like ($\xi_V^h=\xi_u^h=1$).}
\label{Fig:decHpgammagamma}
\end{figure}

\begin{figure*}[tb]
\centering
\includegraphics[width=0.95\textwidth]{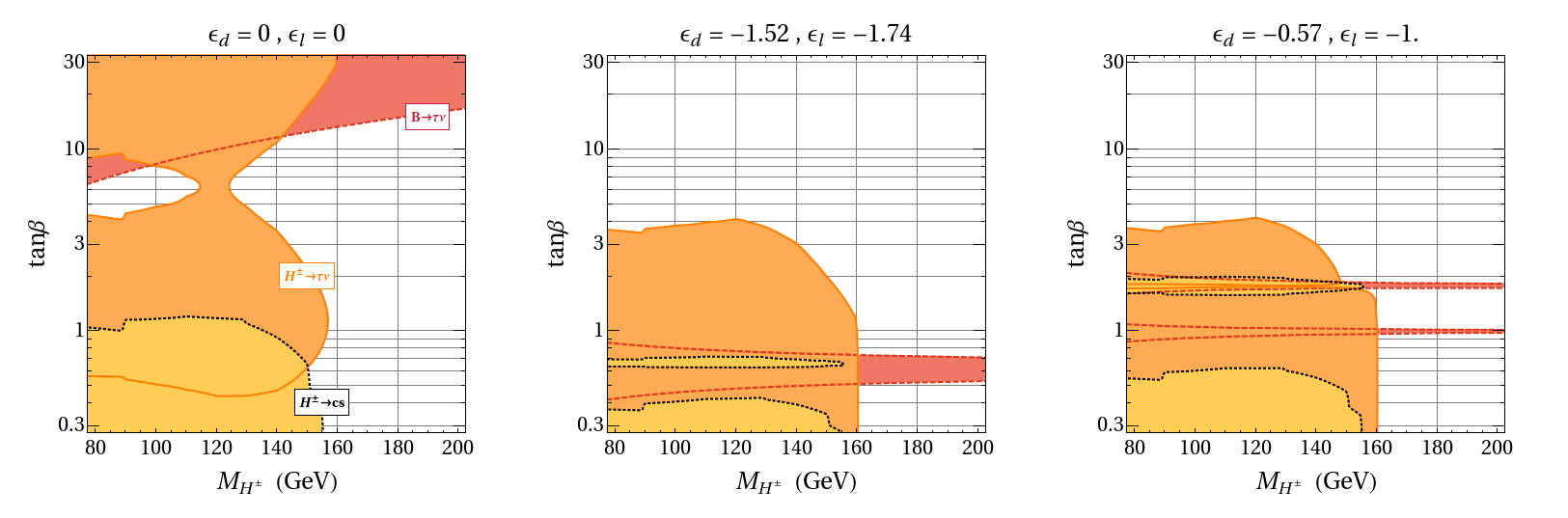}
\caption{Constraints in the $M_{H^\pm}$ -- $\tan\beta$ plane from direct searches for charged Higgs bosons in top decays (the orange region/solid contour corresponds to the $H \to \tau\nu$ final state, the yellow region/dotted contour corresponds to the  $H \to cs$ final state) and from measurements of the $B \to \tau\nu$ branching ratio (red region/dashed contour). All constraints are 95\% C.L.}
\label{fig:MHpm_tanb}
\end{figure*}

A well-known additional non-standard effect in the phenomenology of the light Higgs boson can arise from its interaction with a light charged Higgs that can contribute at 1-loop to the decay $h \to \gamma\gamma$ (for studies of loop induced corrections to the $h \to \gamma\gamma$ rate in new physics models see e.g.~\cite{Blum:2012kn,Carena:2011aa,Batell:2011pz,Arhrib:2011vc,Carena:2012gp,Akeroyd:2012ms,Carena:2012xa,An:2012vp,Joglekar:2012vc,ArkaniHamed:2012kq,Almeida:2012bq,Giudice:2012pf,SchmidtHoberg:2012yy,Davoudiasl:2012ig,Batell:2012ca}. The main contributions to the partial width of the Higgs into two photons are then coming from $W$ loops, top loops, and charged Higgs loops 
\begin{eqnarray} \label{eq:Gamma_hgg}
\Gamma(h\to\gamma\gamma) &\simeq& \frac{\alpha^2m_h^3}{256\pi^3} \frac{1}{v^2} \times \\ \nonumber 
\bigg| \xi_V^h A_1(x_W) 
&+& N_c Q_t^2 \xi_u^h A_{1/2}(x_t) 
+ \frac{\lambda_{h H^\pm H^\pm} v^2}{2M_{H^\pm}^2}A_0(x_{H^\pm})\bigg|^2,
\end{eqnarray}
where $x_i=4M_i^2/M_h^2$, $N_c=3$, $Q_t=2/3$, and $\lambda_{h H^\pm H^\pm}$ is the coupling of the light Higgs boson with two charged Higgs bosons divided by $v$. 
Finally $A_1$, $A_{1/2}$ and $A_0$ are loop functions of a gauge boson, a fermion, and a scalar, respectively, and are collected in Appendix~\ref{sec:appb}.

As shown in Fig.~\ref{Fig:decHpgammagamma}, sizable corrections to the $h \to \gamma \gamma$ partial width are only possible for very light charged Higgs bosons and for large negative coupling with the lightest Higgs boson. 

First consider the familiar case of the MSSM, where the coupling of the lightest Higgs boson with the charged Higgs is determined by electroweak gauge couplings and is given by
\begin{eqnarray}\nonumber
\left|\lambda_{h H^\pm H^\pm}^{\rm{MSSM}}\right| &=& \left|\frac{g_2^2+g_1^2}{4} s_{\beta+\alpha} c_{2\beta}+\frac{g_2^2}{2} s_{\beta-\alpha}\right| \\ &\lesssim& \frac{g_2^2}{2}  \sim 0.21\,.
\end{eqnarray}
This coupling is too small to give a visible effect on the $h \to \gamma\gamma$ partial width.

On the contrary, in a generic 2HDM there exist regions of parameter space producing a large negative coupling $\lambda_{h H^\pm H^\pm} \sim -3$ and light charged Higgs bosons, while still remaining compatible with constraints from vacuum stability and electroweak precision observables. These regions can produce an enhancement of the $h \to \gamma \gamma$ rate by a factor as large as $1.5$-$2$.

The full expression for $\lambda_{hH^\pm H^\pm}$ in terms of a general 2HDM Higgs potential parameters is given in Appendix~\ref{sec:appa}. However,
it is easier to understand the parameter dependence of $\lambda_{h H^\pm H^\pm}$ and of the charged Higgs mass in the ``almost'' decoupling limit and at large $\tan\beta$. In this regime, using the definition of the scalar potential, Eq.~(\ref{eq:Higgs_potential}), we have 
\beq
\lambda_{h H^\pm H^\pm}\sim (\lambda_3+\lambda_4)\,,
\eeq
where it follows that $\lambda_3,\lambda_4$ must be negative and sizable to have a sizable enhancement of the Higgs to di-photon rate.
Additionally, $\lambda_4<0$ allows a smaller mass for the charged Higgs boson given that 
\beq
M_{H^\pm}^2=M_A^2+(\lambda_4+\lambda_5)\frac{v^2}{2}\,,
\eeq
and thus further increases the $h \to \gamma\gamma$ rate.

Large negative couplings in the general 2HDM potential could lead to vacuum stabilities issues.  Following \cite{Ferreira:2009jb}, vacuum stability at tree-level is satisfied so long as 
\begin{eqnarray}\nonumber
&& \lambda_1,\,\lambda_2>0 \, ,\\\nonumber
&&\frac{\lambda_1+\lambda_2}{2}+\lambda_3+\lambda_5-2|\lambda_6 + \lambda_7|>0 \, ,\,\\
&& \lambda_3>-\sqrt{\lambda_1\lambda_2} \, , \;\; \lambda_3-|\lambda_5|>-\sqrt{\lambda_1\lambda_2} \, ,
\end{eqnarray}
in order to have a potential bounded from below.  A full treatment of vacuum stability requires minimization of the renormalization-group improved potential, but this is beyond the scope of this paper. Large negative $\lambda_i$ couplings could lead to additional minima deeper than the electroweak breaking minimum, but we checked at tree-level that this does not occur for the scenarios considered here. Finally, we have also checked that our Higgs potential parameters satisfy electroweak precision observables.  
We find that $\Delta S \lesssim 0.1$ throughout the parameter space, while
$\Delta T \lesssim 0.1$ so long as either $M_{H^\pm} \sim M_A$ or
$M_{H^\pm} \sim M_H$ (or both).  
We find that a large coupling $\lambda_{h H^\pm H^\pm} \lesssim -3$ is compatible with all constraints, as long as $\lambda_5 \neq 0$.

Very light charged Higgs bosons are subject to several constraints.
Model independent bounds from LEP exclude charged Higgs masses below $\lesssim 80$~GeV~\cite{Heister:2002ev}.
At the Tevatron and the LHC, charged Higgs bosons are searched for in decays of top quarks, with the charged Higgs decaying either to $\tau \nu$ or into two jets. Limits are obtained for the product of the branching ratios BR$(t \to H^\pm b) \times$BR$(H^\pm \to \tau \nu)$ and BR$(t \to H^\pm b) \times$BR$(H^\pm \to jj)$. 
Additional constraints arise from flavor observables, in particular from the branching ratio of the decay $B_u \to \tau \nu$ that is sensitive to the tree-level exchange of a virtual charged Higgs.
In a 2HDM Type II, all these limits are conveniently presented in the $M_{H^\pm}$ -- $\tan\beta$ plane: the only two parameters that the branching ratios depend on.

The left plot of Fig.~\ref{fig:MHpm_tanb} shows a summary of these constraints in the Type II model. The orange solid contour corresponds to a combination of the $t \to b H^\pm \to \tau \nu$ searches at ATLAS~\cite{Aad:2012tj}, CMS~\cite{:2012cw}, and D0~\cite{Abazov:2009aa}. The yellow region inside the dotted contour is excluded by a combination of $t \to b H^\pm \to cs$ searches at ATLAS~\cite{ATLAS-CONF-2011-094}, CDF~\cite{Aaltonen:2009ke}, and D0~\cite{Abazov:2009aa}. 
The red region above the dashed contour is excluded by the latest combination of $B_u \to \tau \nu$ data from BaBar~\cite{Aubert:2008ac,Collaboration:2012ju} and Belle~\cite{Hara:2010dk,Adachi:2012mm}, that is in reasonable agreement with SM expectations.
The bounds from top decays only exist for $M_{H^\pm} \lesssim 160$~GeV, while obviously no such restriction exist for the constraint from $B_u \to \tau \nu$.
We observe that in the Type II model a window around $\tan\beta \sim 6$ and $M_{H^\pm} \sim 100$~GeV cannot be excluded based on current available data. In this region of parameter space, charged Higgs loops can lead to large enhancements of the $h \to \gamma\gamma$ rate.

In the 2HDM Type MFV, the constraints in the $M_{H^\pm}$ -- $\tan\beta$ plane depend strongly on the $\epsilon$ parameters. Generically we find that, large values of $\tan\beta \gtrsim 5$ are mostly unconstrained by current data for $\epsilon_i$ factors of $\mathcal{O}(1)$. Correspondingly, in the 2HDM Type MFV, large regions of parameter space are open where charged Higgs loops can enhance the $h \to \gamma\gamma$ rate significantly. This is illustrated in the center and right plot of Fig.~\ref{fig:MHpm_tanb}, that show again the constraints in the $M_{H^\pm}$-$\tan\beta$ plane, fixing the values for $\epsilon_b$ and $\epsilon_\tau$ to the best fit values in the one light Higgs case (center) and the two light Higgs case (right) as indicated. Note that these values represent simply example scenarios, since, introducing the effects of the charged Higgs in the di-photon rate, the best fit values will change. Low values of $\tan\beta$ generically remain constrained by charged Higgs searches in top decays. Indeed, low values of $\tan\beta$ correspond to sizable $\bar t_R b_L H^+$ couplings and therefore to large $H^+$ production from top decay.

In principle also the loop induced $b \to s \gamma$ decay sets strong constraints on light charged Higgs bosons. For example, in the 2HDM Type II, the bound $M_{H^\pm} \gtrsim 380$~GeV holds~\cite{Misiak:2006zs,Hermann:2012fc}. This bound would rule out visible charged Higgs effects in $h \to \gamma\gamma$. However, going beyond the Type II model, the $b \to s \gamma$ bound depends not only on the charged Higgs mass, but also on the charged Higgs couplings $\xi_u^-$ and $\xi_d^-$~\cite{Jung:2010ik,Jung:2010ab}. Moreover, being a FCNC process, $b \to s \gamma$ is also sensitive to higher order terms in the expansions of the Higgs couplings, Eqs.~(\ref{eq:yutilde}),(\ref{eq:ydtilde}), that are not relevant for Higgs collider phenomenology. Correspondingly, we do not consider constraints from $b \to s \gamma$ here.

\section{Conclusions}

We have explored the detailed Higgs phenomenology of a 2HDM
based on the MFV principle, in which both Higgs doublets couple to 
up-type and down-type fermions. 
The agility of the model permits several possibilities
to explain the current hints of $h \ra \gamma\gamma$ excess
in both the inclusive and exclusive VBF channels, 
without leading to large enhancements in the gauge boson
or fermionic channels.
Simultaneously, the model allows to accommodate a strongly reduced 
$h \to \tau^+ \tau^-$ rate, still having a SM-like $h\to b\bar b$ rate.  

Current Higgs data is well described in two distinct regimes of the 
light Higgs couplings:  One regime is where the coupling 
of the Higgs to top quarks has opposite sign with respect to the 
Higgs - gauge boson couplings, and the second is a 
quasi-decoupling regime. We find that the first regime is ruled out 
by searches for the heavy Higgs boson $H$.  The quasi-decoupling regime 
is the only viable region of parameter space in the 2HDM Type MFV\@. 

In this regime we find the VBF $h \ra \gamma\gamma$ can be enhanced by 
up to a factor of 3 or more above the SM rate, still being consistent with the 
present LHC Higgs data.  This occurs by simultaneously:
reducing $\Gamma_{b\bar{b}}$ substantially through the MFV parameter $\eps_d$;
virtually eliminating $\Gamma_{\tau^+\tau^-} \simeq 0$ through $\eps_\ell$;
reducing the $gg \ra h$ production through a reduction in the coupling 
of the Higgs with top quarks $\xi_t = \cos\alpha/\sin\beta$;
while leaving the coupling to gauge bosons nearly identical to
the SM, $\xi_V^h \simeq 1$.  As we showed in Table~\ref{tab:predictions},
a large enhancement in the VBF di-photon channel has important consequences 
in the phenomenology of the light as well as of the heavy Higgs. An enhanced
VBF $h \ra \gamma\gamma$ rate implies both a suppressed $h \ra b\bar{b}$ rate 
and a lower bound on the production cross section of the heavy Higgs boson $H$. 
In particular an enhancement in the VBF di-photon channel by a factor of 
$2$-$3$ would automatically imply very good prospects for the detection
of the heavy Higgs boson.

The inclusive $h \ra \gamma\gamma$ rate can be enhanced by up to 
a factor of $\sim 1.5$, purely through a suppression of the total 
light Higgs boson width.  Again there are correlations
between an enhanced inclusive $h \ra \gamma\gamma$ rate with
a slightly suppressed $h \ra b\bar{b}$ rate and a slightly
enhanced $h \ra VV$ rate, as shown in Table~\ref{tab:predictions}.
Here it is important to point out that the Type MFV model
provides a better fit to the existing (combined) data compared to 
a 2HDM Type II, mainly because the prediction for $h \ra VV$ rates 
can be lowered slightly relative to a Type II 2HDM\@.

In addition to the width effects, 
we also showed that a light charged Higgs boson can lead to significant
loop-induced enhancements in the decay rate $h \ra \gamma\gamma$,
up to a factor of 2 relative to the SM rate, as shown in 
Fig.~\ref{Fig:decHpgammagamma}.
For a given charged Higgs mass and coupling $\lambda_{h H^\pm H^\pm}$, 
the contribution is otherwise the same between the 2HDM Type MFV
versus 2HDM Types I-IV\@.  However, there is a considerably wider
range of ($M_{H^\pm}$, $\lambda_{h H^\pm H^\pm}$) that is 
permitted in the 2HDM Type MFV\@.  When the charged Higgs is
light enough for $t \ra H^+ b$ to be present, much of the parameter space 
of a 2HDM Type II model is ruled out by constraints on this
rare top decay mode.  In addition, $B \ra \tau\nu_\tau$ also
rules out a large swath of parameter space at larger 
$\tan\beta \gtrsim 5$-$20$ for charged Higgs masses between
$80$-$200$~GeV\@.  The 2HDM Type MFV is far less restricted by 
these constraints.  
Generally, for $\eps_{d,\ell} = \mathcal{O}(1)$,
the region above $\tan\beta \gtrsim 3$ is fully allowed.
Moreover, the usual difficulties of accommodating such a light
charged Higgs boson from the constraints on $b \ra s\gamma$ 
can be mitigated given that this flavor-changing process
is also sensitive to higher order terms in the MFV expansion for the 
Yukawa couplings shown in Eqs.~(\ref{eq:yutilde}),(\ref{eq:ydtilde}), 
that were not relevant for Higgs collider phenomenology. 

A quasi-decoupling regime provides the best fit the present 
125 GeV Higgs data.  The deviations from the exact decoupling limit, 
that seem to be hinted by the present data, imply
excellent prospects for observing the heavier Higgs $H$ in 
channels typically used to search for a heavy SM-like Higgs boson.

This we showed in Fig.~\ref{fig:predictions_H}, where 
we find that, corresponding to scenarios that are good fits to the present
125 GeV Higgs data, the existing SM-like searches are 
typically sensitive to 
the heavier Higgs up to about $M_H \lesssim 350$~GeV\@.
In addition, the search for $H \ra b\bar{b}$ can in some cases
provide stronger constraints than $H \ra \tau^+\tau^-$, 
due to the enhanced coupling of $H$ to $b$-quarks.
Above this value for the heavier Higgs mass, the mode
$H \ra t\bar{t}$ opens up, and can have a large branching fraction
that is only weakly constrained by present data.  This also 
provides an exciting opportunity -- the possibility of searching
for the heavier Higgs of a 2HDM Type MFV through a 
$t$-$\bar{t}$ resonance.  

The properties of the heavy Higgs boson typically differ substantially 
from the heavy Higgs bosons of the MSSM\@.
The overwhelmingly dominant decay modes throughout (essentially) 
the full mass range are $H \to b\bar{b}$ and $H\to \tau^+\tau^-$, 
where dedicated searches exist. We emphasize that in order to probe 
the heavy scalar of the 2HDM Type MFV, and in general to find or rule out 
general 2HDMs, it is essential to continue the Higgs searches across 
all search channels in particular in SM-Higgs channels in mass regions 
where a SM-like Higgs is already ruled out.

Finally, we showed that both $h$ and $H$ could be simultaneously light, 
which can provide qualitative differences in the SM-like Higgs rates
that have comparatively low mass resolution.  This means that, 
when $h$ and $H$ are within $\simeq 10$~GeV of one another, 
some rates can be added, including $h+H \ra WW$, $h+H \ra b\bar{b}$, and 
$h+H \ra \tau^+\tau^-$.  This ``two light Higgs boson'' case means 
the total rate into $b\bar{b}$ could exceed the SM value, 
while \emph{simultaneously} having a sizable increase in the 
inclusive and exclusive VBF rates of $h \ra \gamma\gamma$.  
This is generally not possible in the case of just one light Higgs boson.
In addition, in this scenario, there is a small rate of $H$ into the 
high resolution 
channels $H \ra \gamma\gamma$ and $H \ra ZZ$.  The continued 
exploration for resonances in these SM-like Higgs search channels,
but at suppressed rates, would provide an outstanding opportunity
to find or rule out this intriguing two light Higgs boson framework.

\section*{Acknowledgments}

We thank Howard Haber, 
Adam Martin, and Felix Yu for discussions. 
We thank the Aspen Center for Physics for warm hospitality 
where part of this work was completed. 
The Aspen Center for Physics is supported by the 
National Science Foundation Grant No. PHY-1066293.
GDK was supported 
in part by the US Department of Energy under contract number 
DE-FG02-96ER40969.
Fermilab is operated by Fermi Research Alliance, LLC under 
Contract No. DE-AC02-07CH11359 with the United States Department of Energy.

\appendix

\section{Two Higgs Doublet Model Potential}
\label{sec:appa}

The most general renormalizable Higgs potential of a 2HDM can be written as
\begin{eqnarray}\label{eq:Higgs_potential}
V &=& m_{H_1}^2H_1^\dagger H_1 + m_{H_2}^2 H_2^\dagger H_2  \\
   && + \frac{\lambda_1}{2} (H_1^\dagger H_1)^2 + \frac{\lambda_2}{2} (H_2^\dagger H_2)^2 \nonumber \\
   && + \lambda_3 (H_1^\dagger H_1)(H_2^\dagger H_2) + \lambda_4 (H_2^\dagger H_1)(H_1^\dagger H_2) \nonumber \\
   && + \Big( B\mu (H_2 H_1) + \frac{\lambda_5}{2} (H_2 H_1)^2 \nonumber \\
   && - \lambda_6 (H_2 H_1)H_1^\dagger H_1 - \lambda_7 (H_2 H_1)H_2^\dagger H_2 ~+~ h.c. \Big)~, \nonumber
\end{eqnarray}
with $(H_2 H_1) = H_2^+ H_1^- - H_2^0 H_1^0$. The parameters $B\mu$, $\lambda_5$, $\lambda_6$ and $\lambda_7$ are in general complex. However in our study we consider all the parameters real. 

Assuming CP conservation, the Higgs fields entering Eq.~(\ref{eq:Higgs_potential}) can be parameterized in the following way
\begin{eqnarray}\label{eq:Higgs_fields}
H_2 &=& \begin{pmatrix} H_2^+ \\ \frac{1}{\sqrt{2}} (v s_\beta + h_2 + i a_2) \end{pmatrix} ~, \nonumber \\
H_1 &=& \begin{pmatrix} \frac{1}{\sqrt{2}} (v c_\beta + h_1 + i a_1) \\ H_1^- \end{pmatrix}~,
\end{eqnarray}
and the real and imaginary parts of the Higgs fields do not mix. The physical spectrum consists of a charged Higgs $H^\pm$, two scalars $h$ and $H$ and a pseudoscalar $A$.
The Goldstone bosons $G^\pm$ and $G$ provide the longitudinal components of the $W$ and $Z$ bosons, respectively
\begin{eqnarray}
\begin{pmatrix} h \\ H \end{pmatrix} &=& \begin{pmatrix} c_\alpha & -s_\alpha \\ s_\alpha & c_\alpha \end{pmatrix} \begin{pmatrix} h_2 \\ h_1 \end{pmatrix} ~, \nonumber \\
\begin{pmatrix} G \\ A \end{pmatrix} &=& \begin{pmatrix} s_\beta & -c_\beta \\ c_\beta & s_\beta \end{pmatrix} \begin{pmatrix} a_2 \\ a_1 \end{pmatrix} ~, \nonumber \\
\begin{pmatrix} G^\pm \\ H^\pm \end{pmatrix} &=& \begin{pmatrix} s_\beta & -c_\beta \\ c_\beta & s_\beta \end{pmatrix} \begin{pmatrix} H_2^\pm \\ H_1^\pm \end{pmatrix} ~.
\end{eqnarray}

The coupling of the light scalar $h$ with two charged Higgs bosons can be written in terms of the $\lambda_i$ couplings, $\tan\beta$, and $\alpha$ as
\begin{eqnarray} 
&& \lambda_{h H^\pm H^\pm} = -\lambda_1s_\alpha s_\beta^2 c_\beta+\lambda_2 c_\alpha c_\beta^2 s_\beta \\ \nonumber
&&~~ +\lambda_3(c_\alpha s_\beta^3- s_\alpha c_\beta^3) + \lambda_4 s_{\beta-\alpha} +\lambda_5 s_\beta c_\beta c_{\alpha+\beta} \\ \nonumber 
&&~~ + \lambda_6(c_{\alpha+\beta} s_\beta^2 + 2 s_\beta s_\alpha c_\beta^2 ) + \lambda_7(c_{\alpha+\beta} c_\beta^2 + 2 c_\beta c_\alpha s_\beta^2 ) ~.
\end{eqnarray}

\section{One Loop Functions}
\label{sec:appb}

The loop functions appearing in the expression for the $h \to \gamma\gamma$ partial width, Eq.~(\ref{eq:Gamma_hgg}), are given by
\begin{eqnarray}
A_1(x)&=& -x^2\left[2x^{-2}+3x^{-1}+3(2x^{-1}-1)f(x^{-1})\right] ~, \nonumber\\
\label{eq:loop2}
A_{1/2}(x) &=& 2  \, x^2 \left[x^{-1}+ (x^{-1}-1)f(x^{-1})\right] ~, \nonumber\\
 \label{eq:loop3}
A_0(x) &=& -x^2 \left[x^{-1}-f(x^{-1})\right] ~, \nonumber
\end{eqnarray}
with $f(z) = \arcsin^2(\sqrt{z})$ for $z < 1$, which is the case that is relevant for us.



\end{document}